\documentclass[useAMS,usenatbib,usegraphicx]{mn2e}
\usepackage{psfig}   
\usepackage{graphicx}
\usepackage{subfigure}

\title[Binaries in the field]{Binaries in the field: fossils of the star formation process?}

\author[R.~J.~Parker \& M.~R.~Meyer]{
  Richard J.~Parker$^{1,2}$\thanks{E-mail: R.J.Parker@ljmu.ac.uk} and Michael R.~Meyer$^2$
  \vspace*{0.1cm}\\
$^{1}$Astrophysics Research Institute, Liverpool John Moores University, 146 Brownlow Hill, Liverpool, L3 5RF, UK\\
$^{2}$Institute for Astronomy, ETH Z{\"u}rich, Wolfgang-Pauli-Strasse 27, 8093 Z{\"u}rich, Switzerland}

\begin{document}

\date{}
                             
\pagerange{\pageref{firstpage}--\pageref{lastpage}} \pubyear{2014}

\maketitle

\label{firstpage}

\begin{abstract}
Recent observations of binary stars in the Galactic field show that the binary fraction and the mean orbital separation both decrease as a function of decreasing primary mass. We present 
$N$-body simulations of the effects of dynamical evolution in star-forming regions on primordial binary stars to determine whether these observed trends can be explained by 
the dynamical processing of a common binary population. We find that dynamical processing of a binary population with an initial binary fraction of unity and an initial excess of intermediate/wide 
separation (100 -- 10$^4$\,au) binaries does not reproduce the observed properties in the field, even in initially dense ($\sim~10^3$\,M$_\odot$\,pc$^{-3}$) star-forming regions.

If instead we adopt a field-like population as the initial conditions, most brown dwarf and M-dwarf binaries are dynamically hard and their overall fractions and separation distributions are unaffected by dynamical evolution. G-dwarf 
and A-star binaries in the field are dynamically intermediate in our simulated dense regions and dynamical processing does destroy some systems with separations $>$100\,au. However, the formation of wide binaries through the dissolution of supervirial regions 
is a strong function of primary mass, and the wide G-dwarf and A-star binaries that are destroyed by dynamical evolution in subvirial regions are replenished by the formation of binaries in supervirial regions. We therefore suggest that 
the binary population in the field is \emph{indicative} of the primordial binary population in star-forming regions, at least for systems with primary masses in the range 0.02 -- 3.0\,M$_\odot$.
\end{abstract}


\section{Introduction}

Understanding the origin of the Galactic field population is one of the most important outstanding problems in astrophysics. The field is likely a mixture of many types of star-forming region, 
but at present we have little information on the `average' star forming region that populates the field in terms of its mass, density, kinematics, chemical signature and binary properties. 

The initial mass function (IMF) of stars is one potential clue to the dominant star formation event. If the IMF were to vary as a function of environment, then this would place constraints on the star formation event which 
dominantly contributes to the field. However, many studies of star-forming (SF) regions, clusters and associations over the past twenty 
years suggest that the environment in which stars form has little influence on the IMF, which appears to be invariant in Galactic SF regions, and is the same as in the field \citep*[][and references therein]{Bastian10}. 

Binary stars are potentially more of a strong constraint on the origin of the Galactic field than the IMF. The seminal paper by \citet{Duquennoy91} found that the multiplicity fraction (hereafter `binary fraction') of Solar-type G-dwarf stars (primary masses in the range 0.8 -- 1.2\,M$_\odot$) 
to be $f_{\rm bin} = 0.58$, where
\begin{equation}
f_{\rm bin} = \frac{B + T + ...}{S + B + T + ...},
\end{equation}
and $S$, $B$ and $T$ are the number of single, binary and triple systems, respectively. These authors also demonstrated that the period distribution can be approximated with a log-normal distribution extending over many orders of magnitude; from spectroscopic (close) 
binary systems with separations $\sim 10^{-3}$\,au to extremely wide  (`common proper motion') systems with separations $\sim 10^5$\,au. In general, the surveys of binary stars are usually sensitive to companions with a mass ratio $q > 0.1$ \citep[e.g.][]{Duchene13b}, where 
\begin{equation}
q = \frac{m_s}{m_p}
\end{equation}
and $m_p$ and $m_s$ are the masses of the primary (usually more massive) and secondary component stars, respectively.

Following the work on Solar-type primaries, \citet{Fischer92} collated the binary statistics for lower-mass M-dwarf stars. Unfortunately, the data were not as complete as for the G-dwarfs, but suggested that the binary fraction of 
M-dwarfs was slightly lower ($f_{\rm bin} = 0.42$) but with a log-normal separation distribution with a similar mean and variance to the G-dwarfs.

In principle, it should be possible to compare the overall fraction, separation distribution (and other orbital parameters) of binaries in SF regions to those in the field. Unfortunately, observations of binaries in SF regions are often limited to a (comparatively) narrow 
separation range \citep[usually 10s -- 1000s au, ][and references therein]{King12b}. Depending on the local density of a SF region, it is usually these systems -- `intermediate' binaries -- which are susceptible to destruction through dynamical encounters \citep{Heggie75,Hills75a,Hills75b}, 
often in a way that is difficult to account for in a simple analytical model \citep{Fregeau06,Parker12b}.

This in turn makes comparing the binary statistics between different regions somewhat difficult. As an example, the Orion Nebula Cluster (ONC) contains no wide ($>$1000\,au) binaries \citep{Scally99}, and in the range 62 -- 620\,au has a binary fraction and separation distribution which is 
consistent with the field \citep{Reipurth07}. On the other hand, the Taurus association appears to contain an excess of wide binaries \citep{Kohler98}, as do several other regions and open clusters \citep[e.g.][]{Patience02,Kohler08}. Do these differences between regions point to different 
star formation outcomes \citep[as suggested by][]{King12b}, or are they merely the result of dynamical evolution of a common primordial population in regions with different densities \citep{Marks12}?

The apparent excess of binary systems with intermediate/wide separations (10s -- 1000s au) in some SF regions led \citet{Kroupa95a} to postulate that the primordial binary fraction could be as high as unity, and that binaries form from a universal initial period distribution \citep{Kroupa95b,Kroupa11}, 
which is modified by dynamical interactions in dense regions and clusters such as the ONC, but not in more sparse regions such as Taurus. This model has been invoked to explain the binary properties in several regions \citep{Marks12}, although the comparison with observations is necessarily limited 
to the observed separation range. \citet{Kroupa95a,Kroupa99} and \citet{Marks12} suggest that the binary fraction ($\sim 0.5$) and log-normal separation distribution in the field result from the dynamical processing of binaries which form from the \citet{Kroupa95b} universal period distribution and a 
binary fraction of unity.

In recent years, several groups of authors have conducted new observations of binary stars in the field, and updated the statistics. \citet{Raghavan10} revisited the Solar-type G-dwarfs in the field and re-affirmed the earlier work by \citet{Duquennoy91}; they found an overall binary fraction of 0.46 and a 
log-normal period/separation distribution with a peak at $\bar{a} = 50$\,au. Most notably, \citet{Bergfors10} showed that the M-dwarfs in the field have a lower binary fraction -- 0.34 -- than that determined by \citet{Fischer92},  and \citet{Janson12} demonstrated that the M-dwarf separation distribution 
peaks at a significantly  lower value than for the G-dwarfs (16\,au instead of 50\,au). 

\begin{figure}
\begin{center}
\rotatebox{270}{\includegraphics[scale=0.33]{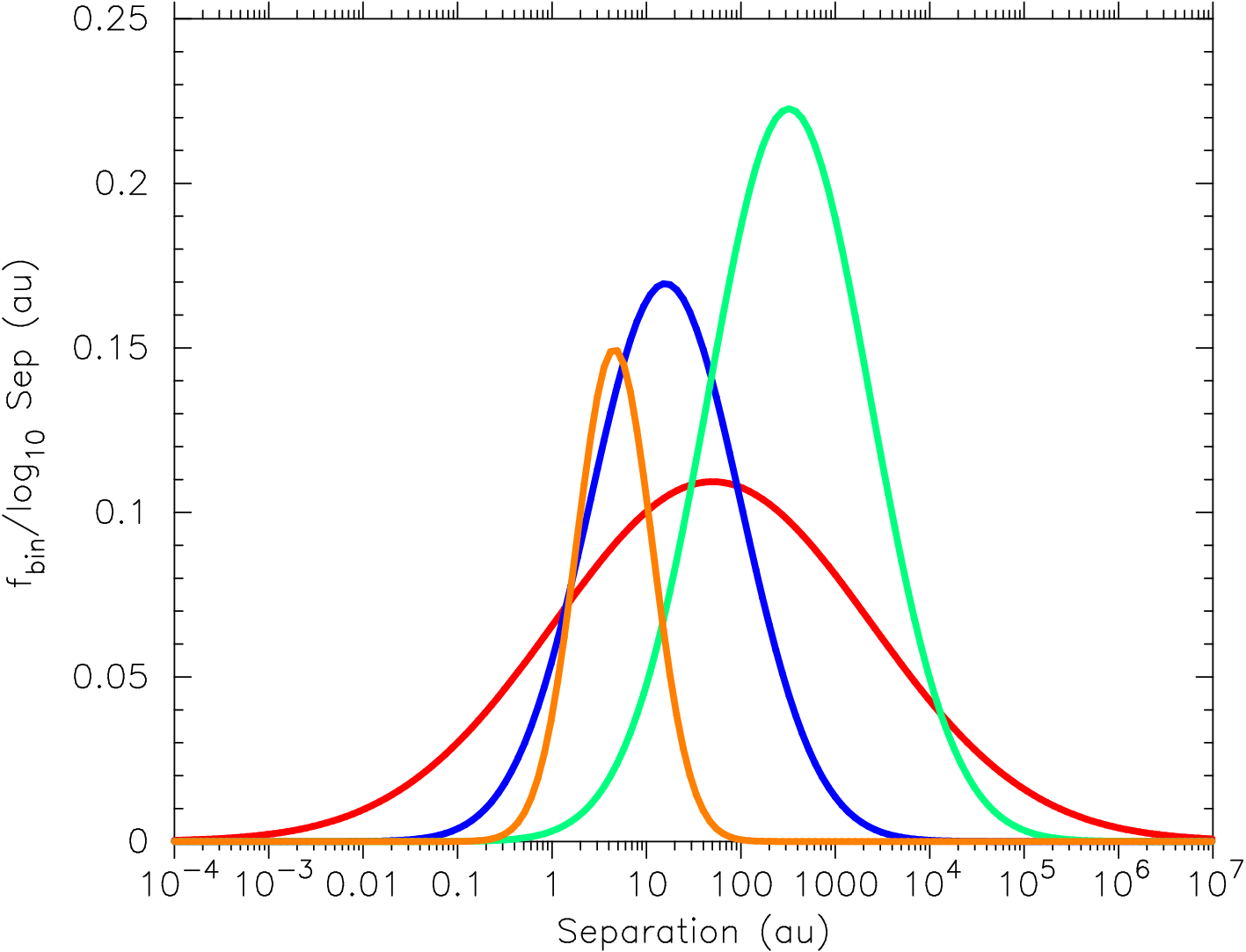}}
\end{center}
\caption[bf]{The log-normal fits to binary separation distributions in the Galactic field. From right to left, the fit to A-star binaries \citep{DeRosa14} is shown by the green line, 
the fit to G-dwarfs \citep{Raghavan10} is shown by the red line, the fit to M-dwarfs is shown by the blue line \citep{Janson12} and the fit to very low mass binaries 
\citep{Burgasser07,Thies07} is shown by the orange line. See Table~\ref{field_props} for details of each log-normal fit. Each distribution is normalised to the binary fraction 
in the field.}
\label{log_normals}
\end{figure}

\begin{figure}
\begin{center}
\rotatebox{270}{\includegraphics[scale=0.33]{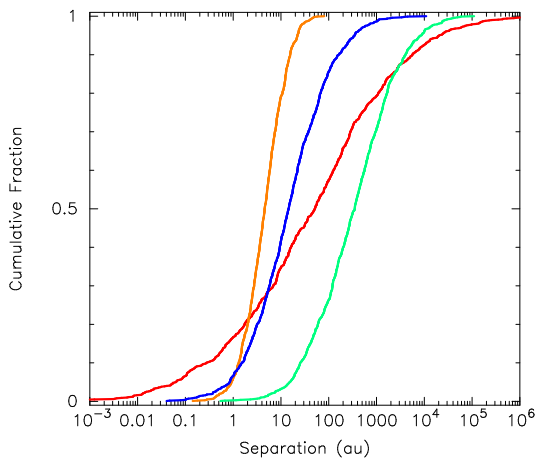}}
\end{center}
\caption[bf]{Cumulative distributions of the log-normal fits to binary separation distributions in the Galactic field. From right to left, the distribution for A-star binaries 
\citep{DeRosa14} is shown by the green line, 
the fit to G-dwarfs \citep{Raghavan10} is shown by the red line, the fit to M-dwarfs is shown by the blue line \citep{Janson12} and the fit to very low mass binaries 
\citep{Burgasser07,Thies07} is shown by the orange line.}
\label{cumulatives}
\end{figure}

Observations of A-star binaries \citep{deRosa12,DeRosa14} show that they have a binary fraction of 0.48, slightly higher than that of G-dwarfs, but their separation distribution peaks at much higher values (389\,au), whereas observations of brown dwarf binaries \citep{Burgasser07} show they have a binary 
fraction of 0.15 and a separation distribution that peaks at much lower values (4.6\,au) than the more massive M-, G- and A-type binaries. In Fig.~\ref{log_normals} we show the log-normal fits to the separation distributions, normalised to the respective binary fractions for the brown dwarfs (orange line), 
M-dwarfs (blue line), G-dwarfs (red line) and A-stars (green line). In Fig.~\ref{cumulatives} we show these separation distributions as cumulative distributions (with the same colour scheme). A summary of these distributions and the parameters used to create them, along with the literature references, are 
provided in Table~\ref{field_props}. 

In addition to the decreasing peak of the separation distribution with decreasing primary mass, the width (i.e.\,\,variance) of the distribution also decreases from G-dwarfs to brown dwarfs, although the width of the A-star distribution is similar to the M-dwarf distribution. \citet{DeRosa14} point out that their observations of A-stars are not sensitive to sub-30\,au binaries, which could imply that the separation distribution for A-stars is wider than observed. \citet{Duchene13b} argue for a double-peaked separation distribution for A-star binaries, and we discuss this possibility and its implications further in Section~\ref{results}.

\begin{table*}
\caption[bf]{Binary properties of systems observed in the Galactic field. We show the spectral type of the primary mass, the main sequence mass range this corresponds to, the binary fraction $f_{\rm bin}$, 
the observed mean separation $\bar{a}$, and the mean (${\rm log}\,\bar{a}$) and variance ($\sigma_{{\rm log}\,\bar{a}}$) of the log-normal fits to these observed separation distributions.}
\begin{center}
\begin{tabular}{|c|c|c|c|c|c|c|}
\hline 
Type & Primary mass & $f_{\rm bin}$ & $\bar{a}$ & ${\rm log}\,\bar{a}$ & $\sigma_{{\rm log}\,\bar{a}}$ & Ref. \\
\hline
A & $1.5 < m/$M$_\odot \leq 3.0$ & 0.48 & 389\,au & 2.59 & 0.79 & \citet{DeRosa14} \\
\hline
G-dwarf & $0.8 < m/$M$_\odot \leq 1.2$ & 0.46 & 50\,au & 1.70 & 1.68 & \citet{Raghavan10} \\
\hline
M-dwarf & $0.08 < m/$M$_\odot \leq 0.45$ & 0.34 & 16\,au & 1.20 & 0.80 & \citet{Bergfors10,Janson12} \\
\hline
Brown dwarf &  $0.02 < m/$M$_\odot \leq 0.08$ & 0.15 & 4.6\,au & 0.66 & 0.4 & \citet{Burgasser07,Thies07} \\
\hline
\end{tabular}
\end{center}
\label{field_props}
\end{table*}

In summary, the observed binary fraction, and the average separation, appear to decrease as a function of decreasing primary mass in the Galactic field. In this paper, we investigate the extent to which these differences 
are due to the effects of dynamical evolution on a single common primordial binary population (i.e.\,\,a binary fraction of unity and the \citet{Kroupa95b,Kroupa11} initial period distribution). We describe the method for setting up 
dense SF regions and binaries in our $N$-body simulations in Section~\ref{method}, we describe the results in Section~\ref{results}, we provide a discussion in Section~\ref{discuss} and we conclude in Section~\ref{conclude}.

\section{Method}
\label{method}

In this section, we describe the method used to set up and run our numerical simulations of the evolution of binary population in star forming regions.

\subsection{Star forming region set up}

Observations of many young star forming regions suggest that stars form in filamentary distributions \citep{Andre10,Arzoumanian11}, which leads to a hierarchical, or self-similar substructured spatial 
distribution of stars \citep[e.g.][]{Cartwright04,Schmeja06,Gouliermis14}. A convenient way of creating substructure as the initial conditions for $N$-body simulations is to use a fractal distribution \citep{Goodwin04a}, 
where the degree of substructure is described by just one number, the fractal dimension, $D$. 

In this paper, we use very fractal ($D = 1.6$ in three dimensions) distributions as the initial conditions -- although both observations \citep{Cartwright04,Schmeja08,Sanchez09,Gouliermis14} and simulations of star 
formation \citep{Schmeja06,Girichidis12,Dale12a,Dale13} show that higher fractal dimensions (less substructure) are possible. The fractals are set up so that local velocities of stars are correlated \citep{Goodwin04a} 
but distant stars can have very different velocities. For full details of the construction of the fractals, we refer the reader to \citet{Goodwin04a} and \citet{Parker14b}.

The fractals each contain 1500 stars and have a radius $r_F  = 1$\,pc, which leads to local densities of $\rho_{\rm local} \sim 10^3$\,M$_\odot$\,pc$^{-3}$. We adopt two different initial dynamical states for our simulated star forming regions. 
In two sets of simulations we scale the velocities to be subvirial ($\alpha_{\rm vir} = 0.3$, where the virial ratio $\alpha_{\rm vir} =  T/|\Omega|$; $T$ and $|\Omega|$ are the total kinetic energy and total potential energy of the stars, respectively).
This leads to the erasure of substructure within several crossing times, and the region collapses to form a centrally concentrated, bound star cluster. When the substructure has these high initial densities, virtually all of the dynamical destruction  
of binaries occurs before the subsequent collapse and formation of a cluster \citep*{Parker11c}.

In a third set of simulations, we scale the velocities to be supervirial, $\alpha_{\rm vir} = 1.5$. These regions expand, but retain some substructure and can result in the formation of an association-like complex, or a binary cluster \citep{Parker14b}. The initial 
density of the substructure in both our simulated subiviral and supervirial regions is significantly more dense than the majority of nearby star-forming regions \citep{Bressert10}, which will likely contribute to the Galactic field. This is to allow for the possibility 
that the field may be the sum of clusters that were more dense at earlier epochs \citep{Longmore14}, which would imply that the field binaries are more likely to have been dynamically processed, or that the local star-forming regions in \citet{Bressert10} are somehow not representative of all regions.  

We draw \emph{primary} stellar masses from the analytic determination of the IMF by \citet{Maschberger13} which has a probability density function 
of the form:
\begin{equation}
p(m) \propto \left(\frac{m}{\mu}\right)^{-\alpha}\left(1 + \left(\frac{m}{\mu}\right)^{1 - \alpha}\right)^{-\beta}
\label{imf}.
\end{equation}
Eq.~\ref{imf} essentially combines the log-normal approximation for the IMF derived by \citet{Chabrier03,Chabrier05} with the \citet{Salpeter55} power-law slope for stars with mass $>$1\,M$_\odot$. Here, 
$\mu = 0.2$\,M$_\odot$ is the average stellar mass, $\alpha = 2.3$ is the Salpeter power-law exponent for higher mass stars, and $\beta = 1.4$ is the power-law exponent to describe the slope of the 
IMF for low-mass objects \citep[which also deviates from the log-normal form;][]{Bastian10}. Finally, we sample from this IMF within the mass range $m_{\rm low} = 0.01$\,M$_\odot$ to $m_{\rm up} = 50$\,M$_\odot$.

\subsection{Binary populations} 

We utilise two separate binary populations in our simulations. In one set of simulations all binaries form from a `common' primordial population, i.e.\,\,the binary fraction is unity (everything forms in a binary) and the semi-major axes are 
drawn from the same initial distribution. In this scenario, we are testing the hypothesis that the observed decrease in binary fraction and mean separation as a function of primary mass is due to the dynamical processing of a common population, and that systems with 
lower primary masses (and therefore lower binding energy on average) are simply more susceptible to destruction. In the remaining simulations, we use set the binary fractions and semi-major axis distributions as a function of primary mass, as observed in the Galactic field.

In all cases, secondary masses are drawn from a flat mass ratio ($q$) distribution, as observed in the Galactic field and most star-forming regions \citep{Metchev09,Reggiani11a,Reggiani13}. Dynamical evolution is not expected to alter the shape of 
the mass ratio distribution \citep{Parker13b}.  

In all cases, orbital eccentricities are drawn from a flat distribution, as observed for the G-dwarf field binaries \citep{Raghavan10} and also M-dwarfs \citep{Abt06} -- the initial eccentricity distribution (if different from the field) from the star formation process remains unconstrained by observations 
\citep{Duchene13b}.

\subsubsection{Common separation distribution}

In a series of pioneering papers, \citet{Kroupa95a,Kroupa95b} studied the dynamical evolution of binary populations in star clusters and suggested that the primordial binary fraction should be higher than observed in the field, and that an excess of intermediate--wide 
binaries (with separations $>1000$\,au) was necessary to explain the apparent excess of wide binaries the Taurus association compared to the field distributions for G- and M-dwarfs \citep{Duquennoy91,Fischer92}.

\citet{Kroupa95a} suggested a primordial binary population with $f_{\rm bin} = 1$ and a period distribution of the following form:
\begin{equation}
f\left({\rm log_{10}}P\right) = \eta\frac{{\rm log_{10}}P - {\rm log_{10}}P_{\rm min}}{\delta + \left({\rm log_{10}}P - {\rm log_{10}}P_{\rm min}\right)^2},
\end{equation}
where ${\rm log_{10}}P_{\rm min}$ is the logarithm of the minimum period in days and ${\rm log_{10}}P_{\rm min} = 0$. $\eta = 3.5$ and $\delta = 100$ are the numerical 
constants adopted by \citet{Kroupa95a} and \citet{Kroupa11} to fit the observed pre-main sequence distributions.

However, drawing binary periods from a particular distribution and then converting to separation means that if $P$ is constant, $a$ is proportional to $m_1 + m_2$, the binary system mass. So for a similar orbital period, 
an M-dwarf binary will have a smaller semi-major axis than a G-dwarf binary -- and hence is `harder' and could be less likely to be destroyed. To avoid this, we approximate the \citet{Kroupa95a} period distribution in terms of 
semi-major axes as:
\begin{equation}
f\left({\rm log_{10}}a\right) = \eta\frac{{\rm log_{10}}a - {\rm log_{10}}a_{\rm min}}{\delta + \left({\rm log_{10}}a - {\rm log_{10}}a_{\rm min}\right)^2},
\label{coma}
\end{equation}
where ${\rm log_{10}}a$ is the logarithm of the semi-major axis in au and ${\rm log_{10}}a_{\rm min} = -2$ ($a_{\rm min} = 0.01$\,au). The constants are now $\eta = 5.25$ and $\delta = 77$. This then avoids the small differences 
in separation distribution as a function of primary mass when using a common period distribution (although we ran a test simulation and found that the differences to the results are minimal -- see Section~\ref{results}).

\subsubsection{Field separation distributions}

\begin{table}
\caption[bf]{Summary of simulation set-ups. The columns show the simulation suite number, number of stars,  $N_{\rm stars}$, initial virial ratio of the regions, $\alpha_{\rm vir}$, initial binary fraction, $f_{\rm bin}$, 
and the initial semi-major axis distribution, $f(a)$.}
\begin{center}
\begin{tabular}{|c|c|c|c|c|c|}
\hline 
Sim.\,No. & $N_{\rm stars}$ & $\alpha_{\rm vir}$& $f_{\rm bin}$ & $f(a)$ \\
\hline
1 & 1500 & 0.3 & 1.00 & Common (Eq.~\ref{coma}) \\
2 & 1500 & 0.3 & field & field \\
3 & 1500 & 1.5 & field & field \\
\hline
\end{tabular}
\end{center}
\label{cluster_sims}
\end{table}

\begin{table*}
\caption[bf]{A summary of the results for the simulations containing binaries drawn from a common population (identical separation distribution and binary fraction). From left to right, 
the columns are primary component mass-type, the input binary fraction, $f_{\rm bin}$ (init.), the actual binary fraction calculated before dynamical evolution, $f_{\rm bin}$ (0\,Myr), the binary fraction 
calculated after 10\,Myr of dynamical evolution, $f_{\rm bin}$ (10\,Myr), the median separation before dynamical evolution, $\tilde{a}$ (0\,Myr), and the median separation after 10\,Myr of 
dynamical evolution, $\tilde{a}$ (10\,Myr).}
\begin{center}
\begin{tabular}{|c|c|c|c|c|c|}
\hline 
Primary &  $f_{\rm bin}$ (init.) & $f_{\rm bin}$ (0\,Myr) & $f_{\rm bin}$ (10\,Myr) & $\tilde{a}$ (0\,Myr) & $\tilde{a}$ (10\,Myr) \\
\hline
A-star & 1.00  & 0.75 & 0.56 & 47\,au & 18\,au \\
\hline
G-dwarf & 1.00 & 0.73 & 0.55 & 34\,au & 14\,au \\
\hline
M-dwarf & 1.00 & 0.69 & 0.42 & 33\,au & 8.9\,au \\
\hline
Brown dwarf & 1.00 & 0.56 & 0.25 & 30\,au & 5.5\,au \\
\hline
\end{tabular}
\end{center}
\label{summary_cold_common}
\end{table*}

In the remaining simulations, we use the observed binary properties in the Galactic field as the initial conditions for our binary populations. A summary of the differing properties as a function of primary mass is given in Table~\ref{field_props}. 
Systems with a primary mass in the range $0.02 < m/$M$_\odot \leq 0.08$ are brown dwarf binaries, with a corresponding fraction $f_{\rm bin} = 0.15$ and a log-normal semi-major axis distribution with mean ${\rm log}\,\bar{a} = 0.66$ and variance 
$\sigma_{{\rm log}\,\bar{a}} = 0.40$ \citep{Burgasser07,Thies07}. Systems with primary masses in the range $0.08 < m/$M$_\odot \leq 0.45$ are M-dwarf 
binaries, with a fraction $f_{\rm bin} = 0.34$ and a log-normal semi-major axis distribution with mean ${\rm log}\,\bar{a} = 1.20$ and variance $\sigma_{{\rm log}\,\bar{a}} = 0.80$ \citep{Bergfors10,Janson12}. 
Systems with primary masses in the range $0.8 < m/$M$_\odot \leq 1.2$ are G-dwarf binaries with a fraction $f_{\rm bin} = 0.46$ and a log-normal semi-major axis distribution with mean ${\rm log}\,\bar{a} = 1.70$ 
and variance $\sigma_{{\rm log}\,\bar{a}} = 1.68$ \citep{Raghavan10}. Systems with primary masses in the range $1.5 < m/$M$_\odot \leq 3.0$ are A-star binaries with  a fraction $f_{\rm bin} = 0.48$ and a log-normal 
semi-major axis distribution with mean ${\rm log}\,\bar{a} = 2.59$ and variance $\sigma_{{\rm log}\,\bar{a}} = 0.79$ \citep{DeRosa14}. There is also evidence of a bimodal distribution for A-stars, 
with a second peak around  0.01\,au \citep{Duchene13b} -- however, these binaries are unlikely to be altered by dynamical evolution and we do not include them in our simulations.

We also include more massive binaries with $f_{\rm bin} = 1.0$ and an \citet{Opik24} distribution of semi-major axes in the range $0 < a < 50$\,au for all binaries with primary masses greater than 3.0\,M$_\odot$, as suggested by 
the observations of \citet{Sana13}, 
although we do not consider these binaries further in our analysis. Any binaries lying outside the mass ranges discussed (e.g. K-type and F-type primaries) are assigned the same properties as the G-dwarfs. Again, these are not 
considered in the subsequent analysis.  \\

We place binaries or single stars at the position of each system in the fractal distribution and run the simulations for 10\,Myr using the \texttt{kira} integrator in the Starlab package \citep{Zwart99,Zwart01}. 
We do not include stellar evolution in the simulations. A summary of the simulation parameter space is 
given in Table~\ref{cluster_sims}.

\section{Results}
\label{results}

In this Section we describe the effects of dynamical evolution on binaries drawn from a single common population in collapsing star-forming regions (Sect.~\ref{results:common}), binaries drawn from the the field populations in collapsing star-forming regions (Sect.~\ref{results:fieldcollapse}) and binaries drawn from the field populations in expanding star-forming regions (Sect.~\ref{results:warmexpand}). 

We identify binaries using the nearest neighbour algorithm described in \citet{Parker09a} and \citet{Kouwenhoven10}. If two stars are mutual nearest neighbours, and their separation is less than a quarter of the mean separation between stars in the simulation, then we determine whether the two stars are energetically bound. If so, we classify them as a binary system. We note that other methods to identify binaries (and multiple systems) are also utilised in the literature \citep[e.g.][]{Bate09}.

\subsection{Common primordial binary population}
\label{results:common}

In the first set of simulations, we draw all the binary systems from a single, common population. The input binary fraction is unity, and the separation distribution is constructed to mimic the 
pre-main sequence period distribution in \citet{Kroupa95a}\footnote{We also used the original period distribution in one set of simulations and converted periods to separations using the component masses; the results are 
very similar.}. We summarise the results in Table~\ref{summary_cold_common}.


\begin{figure}
\begin{center}
\rotatebox{270}{\includegraphics[scale=0.4]{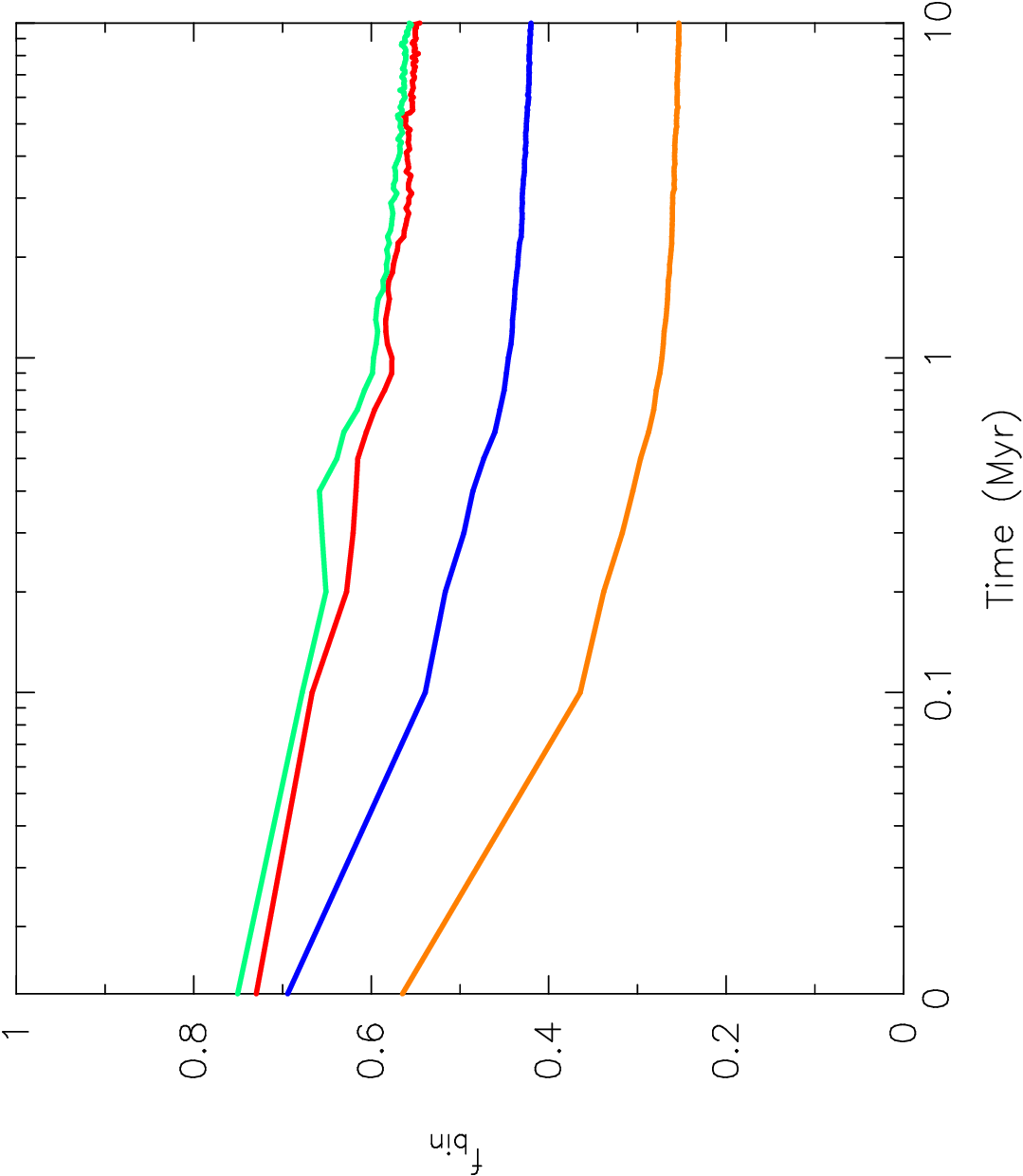}}
\end{center}
\caption[bf]{Evolution of the binary fraction for binaries with properties drawn from a single primordial population in simulated dense star forming regions undergoing cool-collapse. 
The first (top, green) line shows the evolution of the A-star binary fraction; the second (red) line shows the evolution of the G-dwarf binary fraction; the third (blue) line shows the 
evolution of the M-dwarf binary fraction; and the fourth (bottom, orange) line shows the evolution of the brown dwarf binary fraction. }
\label{cold_common_bin_frac}
\end{figure}

\begin{figure*}
  \begin{center}
\setlength{\subfigcapskip}{10pt}
\subfigure[Brown dwarf binaries, 10 Myr]{\label{cold_common_BD_sepdist}\rotatebox{270}{\includegraphics[scale=0.35]{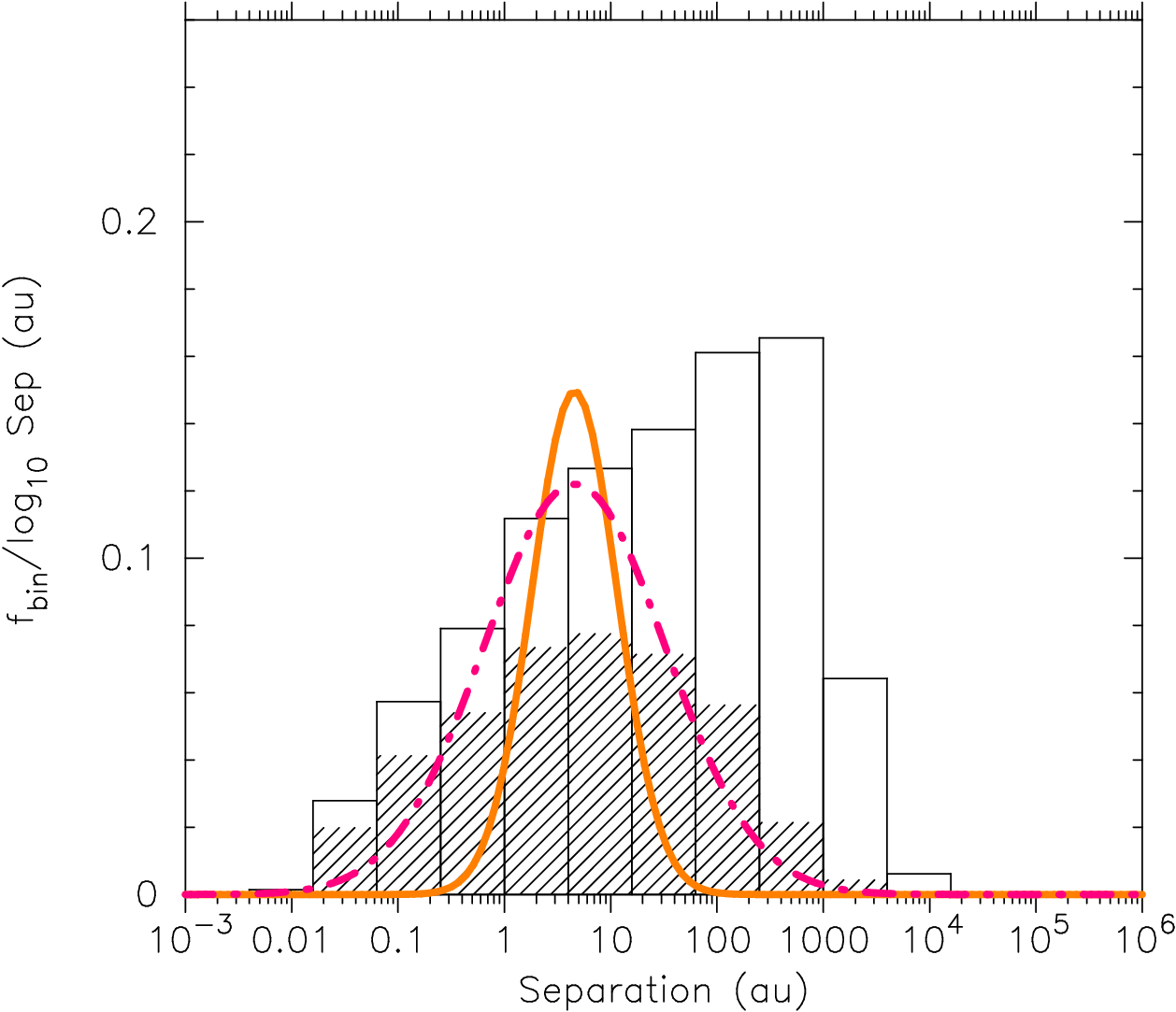}}}
\hspace*{0.8cm}
\subfigure[M-dwarf binaries, 10 Myr]{\label{cold_common_M_sepdist}\rotatebox{270}{\includegraphics[scale=0.35]{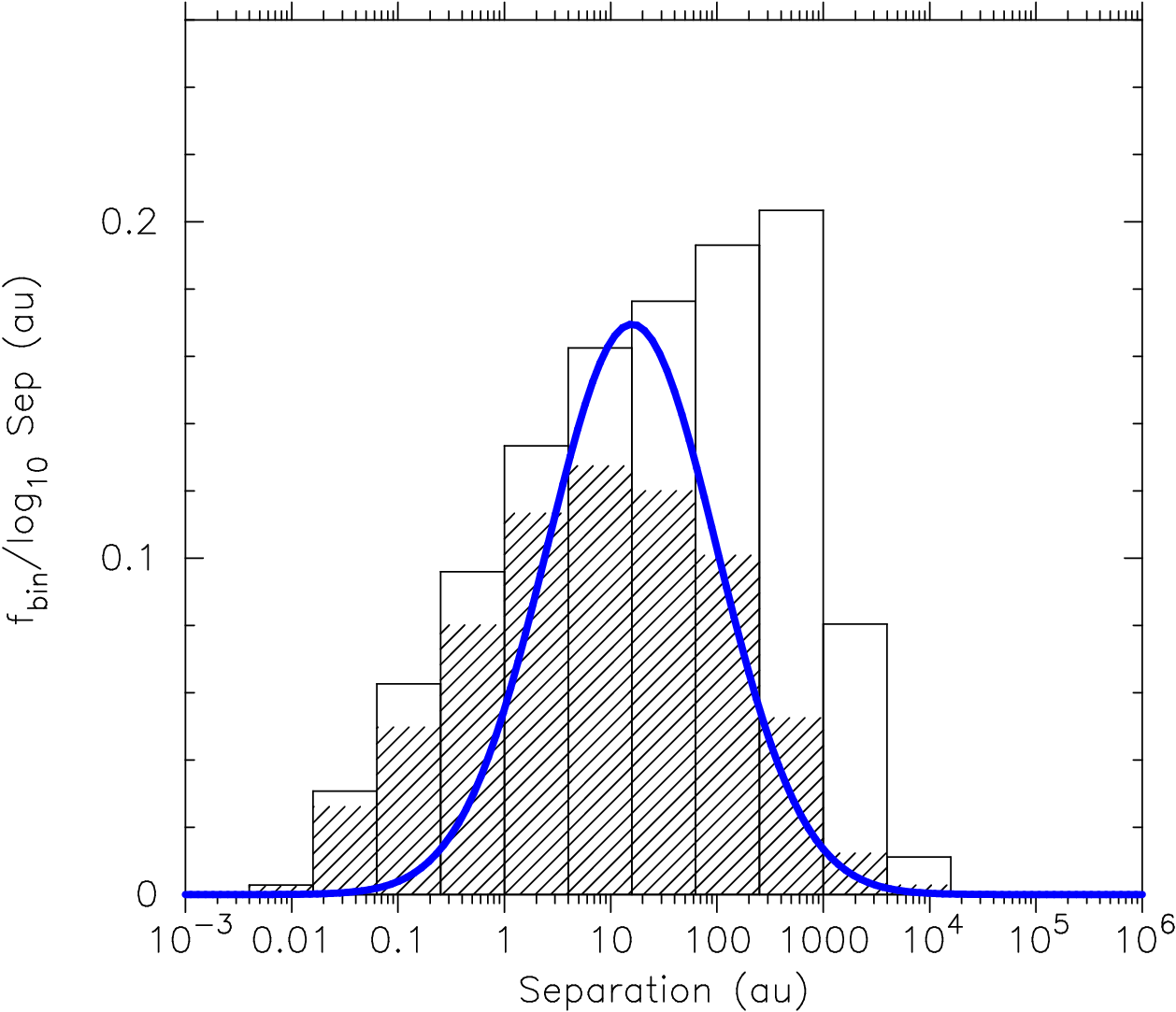}}}
\vspace*{0.25cm}
\subfigure[G-dwarf binaries, 10 Myr]{\label{cold_common_G_sepdist}\rotatebox{270}{\includegraphics[scale=0.35]{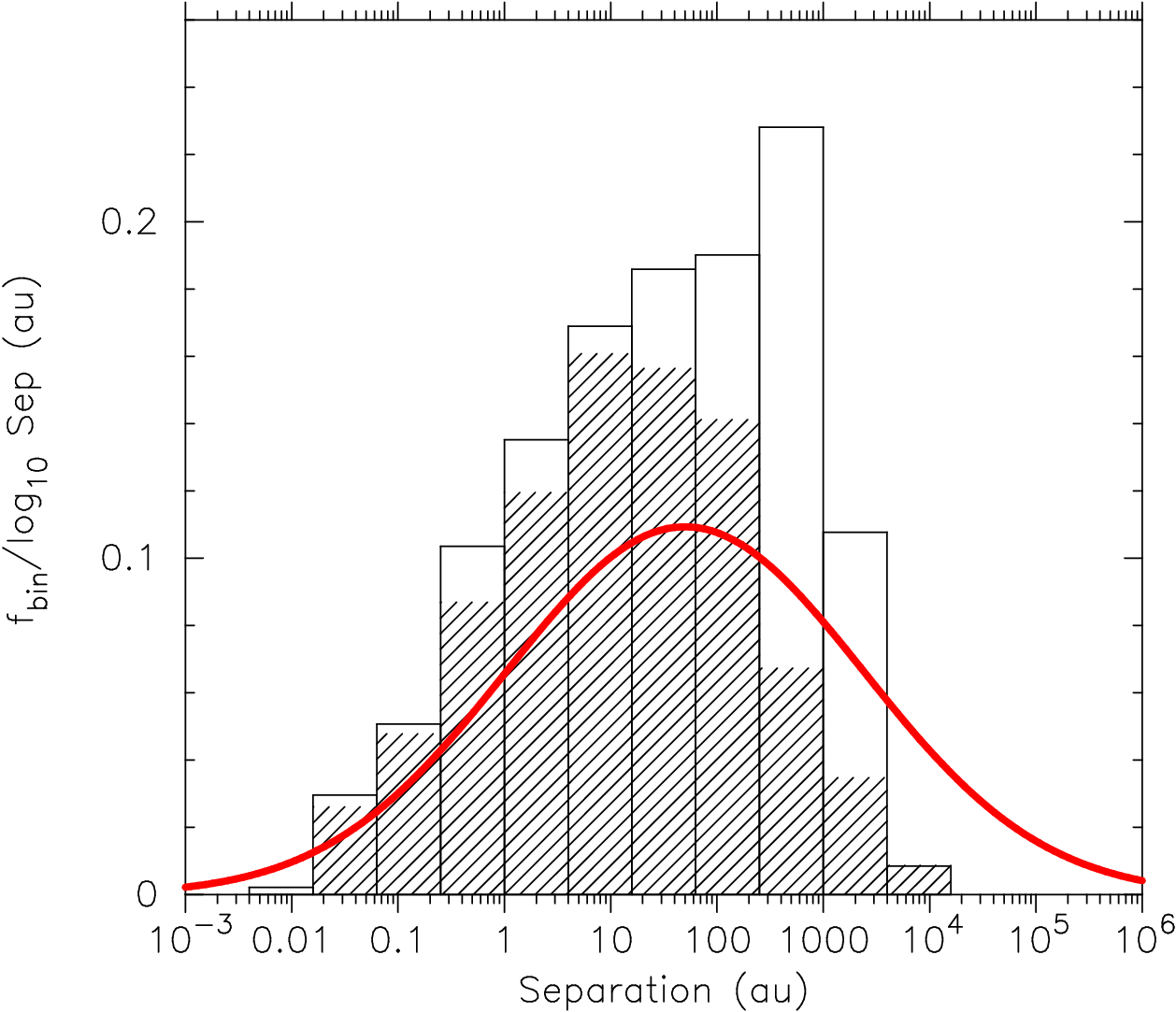}}}
\hspace*{0.8cm}
\subfigure[A-star binaries, 10 Myr]{\label{cold_common_A_sepdist}\rotatebox{270}{\includegraphics[scale=0.35]{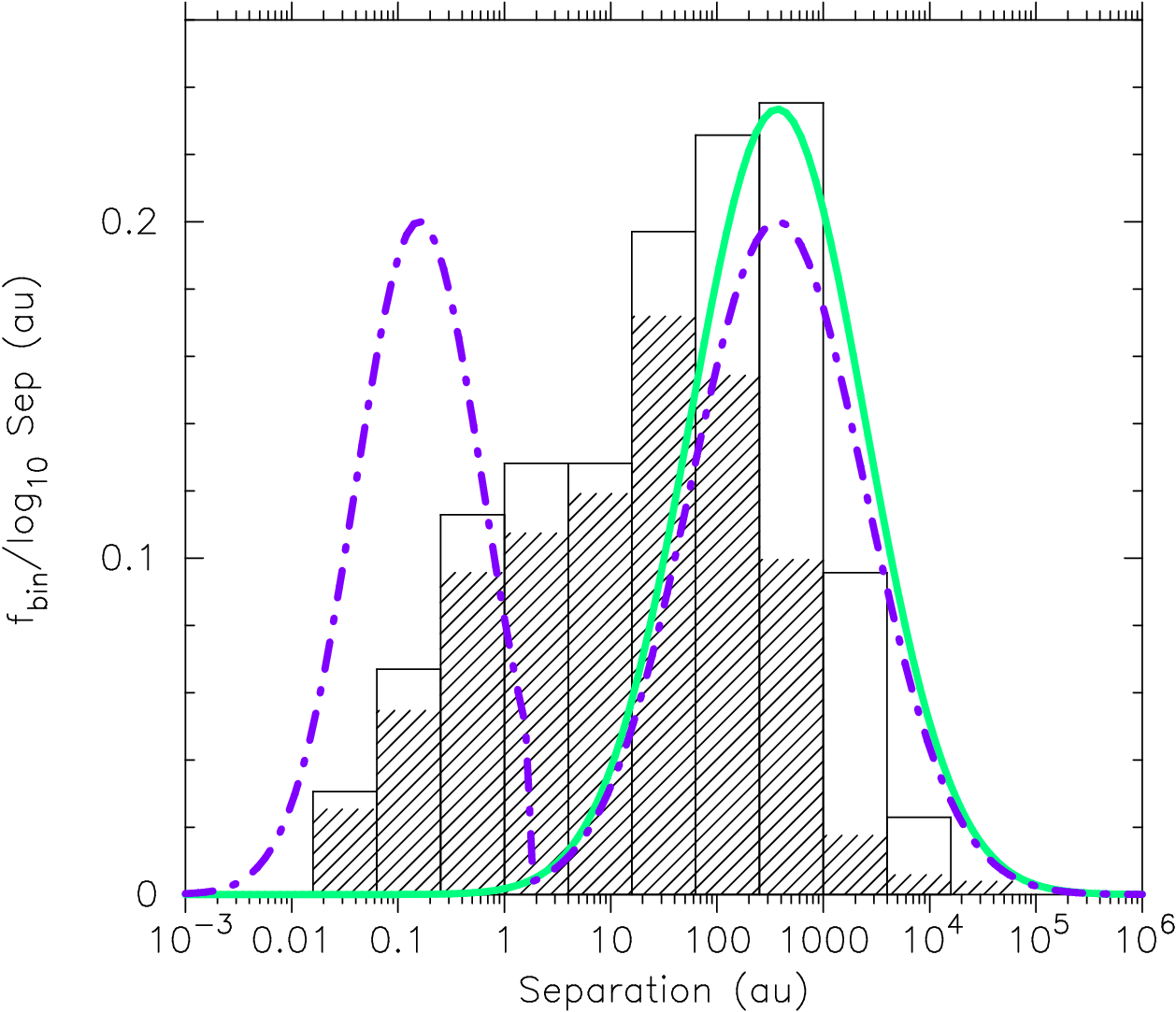}}}
  \end{center}
  \caption[bf]{Evolution of the separation distributions for binaries with properties drawn from a single primordial population in simulated dense star forming regions undergoing cool-collapse. In all panels, 
the open histogram shows the distribution at 0\,Myr (i.e. before dynamical evolution) and the hashed histogram shows the distribution after 10\,Myr. In panel 
(a) we show the evolution of the distribution for brown dwarf (BD) binaries; the log-normal approximation to the data from \citet{Thies07} is shown by the (solid) orange line (normalised to a binary fraction of 0.15), and the log-normal approximation 
to the data assuming `missing' systems \citep{Basri06} is shown by the (dot-dashed) magenta line (normalised to a binary fraction of 0.26). In panel (b) we show the evolution of the distribution for M-dwarf binaries; the log-normal 
approximation to the data by \citet{Janson12} is shown by the (solid) blue line (normalised to a binary fraction of 0.34). In panel (c) we show the evolution of the distribution for G-dwarf binaries; the log-normal approximation to the 
data by \citet{Raghavan10} is shown by the (solid) red line. In panel (d) we show the evolution of the distribution for A-star binaries; the log-normal approximation to the visual binary data by \citet{DeRosa14} is shown by the (solid) green line (normalised 
to a binary fraction of 0.48), and the fit to the bimodal distribution discussed in \citet{Duchene13b} is shown by the (dot-dashed) purple line (normalised to a binary fraction of 0.70).  }
  \label{cold_common_sep_results_dist}
\end{figure*}

\begin{figure*}
  \begin{center}
\setlength{\subfigcapskip}{10pt}
\subfigure[Brown dwarf binaries, 10 Myr]{\label{cold_common_BD_sepcum}\rotatebox{270}{\includegraphics[scale=0.35]{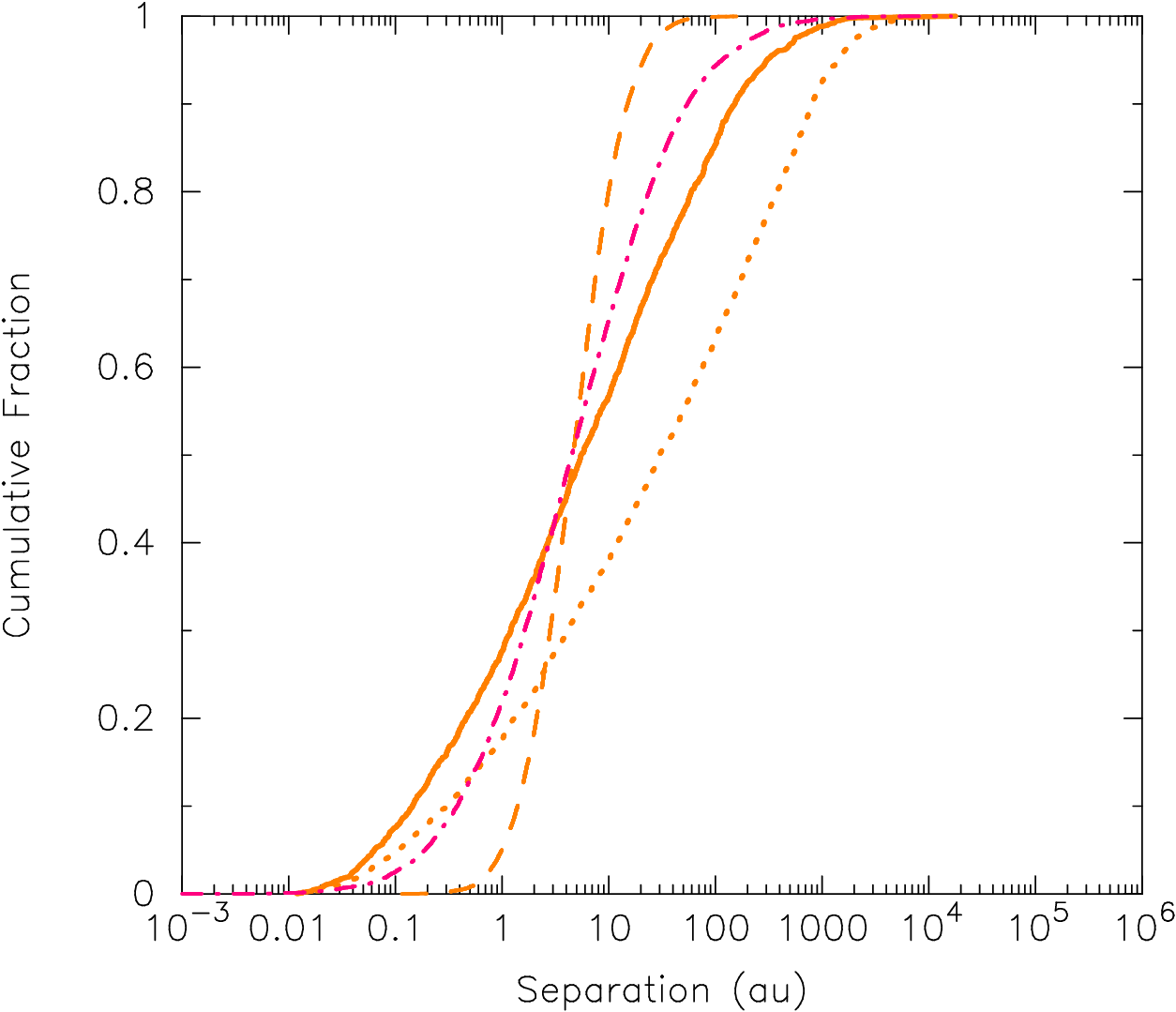}}}
\hspace*{0.8cm}
\subfigure[M-dwarf binaries, 10 Myr]{\label{cold_common_M_sepcum}\rotatebox{270}{\includegraphics[scale=0.35]{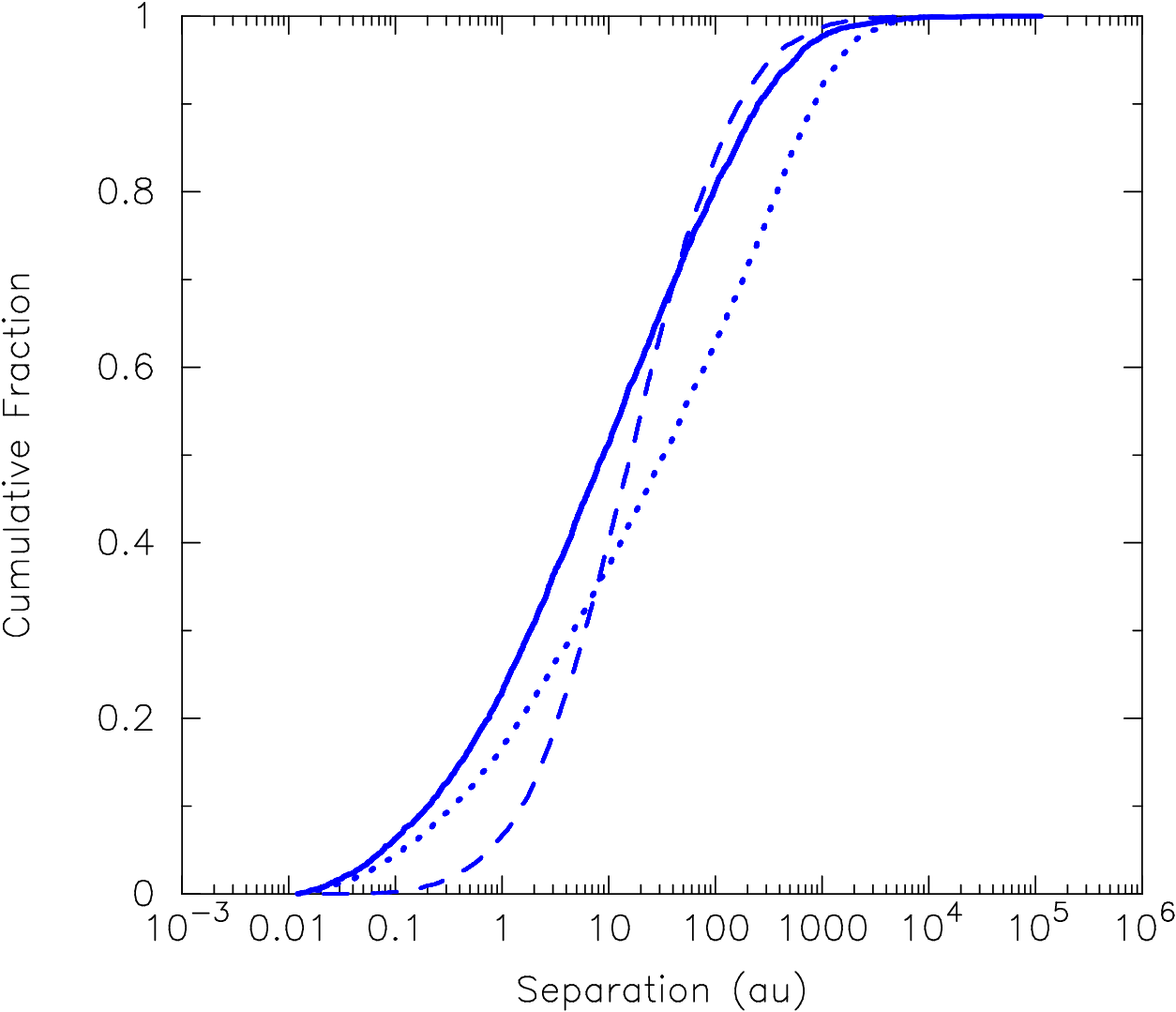}}}
\vspace*{0.25cm}
\subfigure[G-dwarf binaries, 10 Myr]{\label{cold_common_G_sepcum}\rotatebox{270}{\includegraphics[scale=0.35]{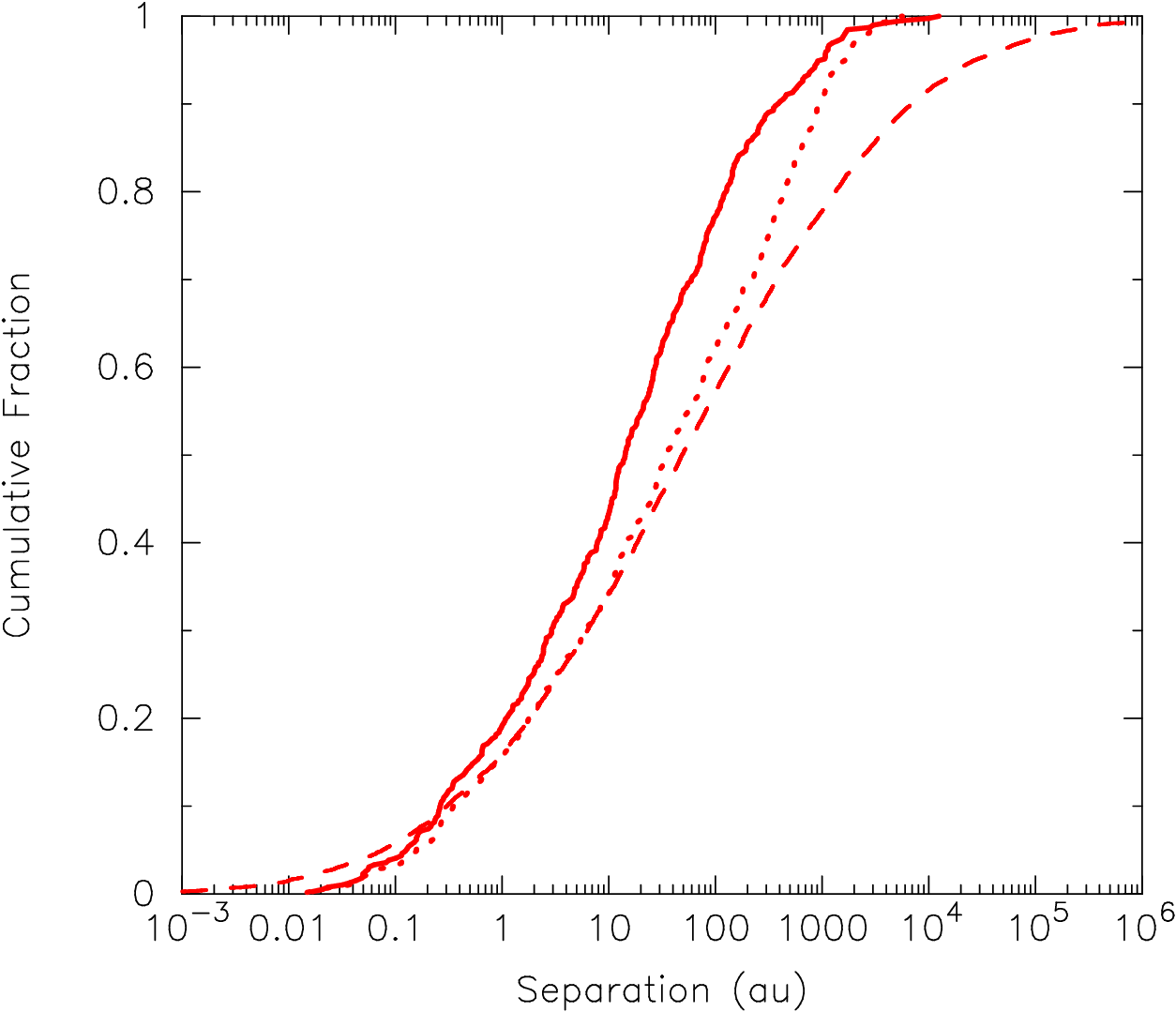}}}
\hspace*{0.8cm}
\subfigure[A-star binaries, 10 Myr]{\label{cold_common_A_sepcum}\rotatebox{270}{\includegraphics[scale=0.35]{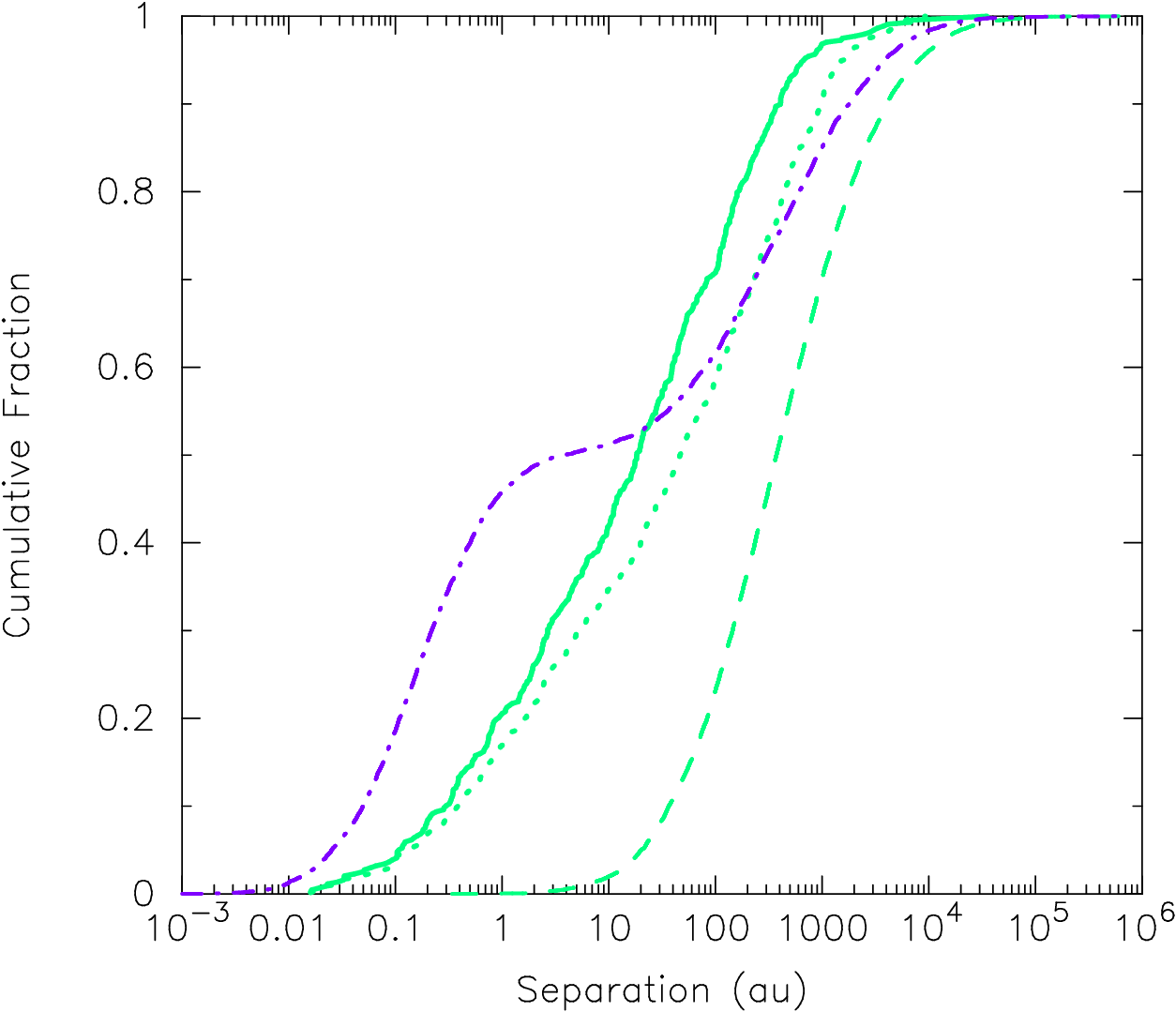}}}
  \end{center}
  \caption[bf]{Evolution of the cumulative separation distributions for binaries with properties drawn from a single primordial population in simulated dense star forming regions undergoing cool-collapse. In all panels, 
the dotted line shows the distribution at 0\,Myr (i.e. before dynamical evolution) and the solid line shows the distribution after 10\,Myr. In all panels the dashed lines show the respective cumulative distributions of the log-normal fits to the data for each primary mass range observed in the Galactic field 
(detailed in Table~\ref{field_props}). In panel 
(a) the cumulative distribution proposed by \citep{Basri06} is shown by the dot-dashed magenta line. In panel (d) the bimodal distribution discussed in \citet{Duchene13b} is shown by the dot-dashed purple line.}
  \label{cold_common_sep_results_cum}
\end{figure*}

As noted in \citet{Parker11c}, the large number of wide ($> 10^4$\,au) systems generated by this distribution precludes them from being physically bound in our locally dense \citep[$\rho_{\rm local} \sim 10^3$\,M$_\odot$\,pc$^{-3}$,][]{Parker14b} 
simulated regions. The binary fractions at 0\,Myr (i.e.\,\,before dynamical evolution), are therefore substantially lower than the `initial' binary fraction of unity (see Fig.~\ref{cold_common_bin_frac}). 

In Fig.~\ref{cold_common_bin_frac} we show the evolution of the binary fraction for our four chosen primary mass ranges. Systems with a brown dwarf primary are shown by the (bottom) orange line, those with an M-dwarf primary are shown by the 
(lower middle) blue line, those with a G-dwarf primary are shown by the (upper middle) red line and the (top) green line is for A-star primaries. We immediately see that fewer low-mass binaries are bound at 0\,Myr compared to high-mass systems (for example, 
the binary fraction of brown dwarf binaries at 0\,Myr is 0.56, compared to 0.75 for A-star binaries). 

The subsequent dynamical evolution (both in the substructure, and when the region collapses to form a spherical cluster) reduces the binary fractions further. After 10\,Myr the binary fraction of brown dwarf binaries 
is 0.25, compared to 0.42 for the M-dwarfs, 0.55 for the G-dwarfs, and 0.56 for the A-stars. At first sight, dynamical evolution of a common binary population appears to result in binary fractions roughly consistent (although a little high) with those observed in the 
Galactic field, and the trend for lower binary fraction with lower primary mass is recovered. 

However, examination of the separation distribution after 10\,Myr suggests that a common primordial binary fraction and separation distribution cannot be the dominant initial conditions for star formation in binaries. In Fig.~\ref{cold_common_sep_results_dist} we show histograms 
of the separation distributions at 0\,Myr (before dynamical evolution; the open histogram in all panels) and at 10\,Myr (the hashed histograms). In Fig.~\ref{cold_common_BD_sepdist} we show the log-normal fit by \citet{Thies07} to the observed separation distribution of very low mass 
binaries presented in \citet{Burgasser07} by the solid orange line, which has a mean separation of $\bar{a} = 4.6$\,au. We also show the log-normal fit by \citet{Basri06}, which accounts for potentially closer \citep{Maxted05} and wider \citep{Bouy06,Dhital11} brown dwarf binaries that remain undiscovered (the dot-dashed magenta line). 
The histograms of the binaries in the simulations are normalised to the respective binary fractions at 0\,Myr (0.56) and 10\,Myr (0.25), whereas the fits to the data are normalised to the observed binary fractions \citep[0.15,][]{Thies07}, or 0.26 in the case of the fit by \citet{Basri06}.

In Fig.~\ref{cold_common_M_sepdist} we show the log-normal fit to the observed separation distribution of M-dwarf binaries by \citet{Janson12}, which has a mean separation of $\bar{a} = 16$\,au, by the solid blue line. This is normalised to the binary fraction of M-dwarfs in the field 
\citep[0.34;][]{Bergfors10}. The open histogram shows the distribution of binary separations at 0\,Myr in the simulations, normalised to the initial binary fraction (0.69), and the hashed histogram shows the distribution at 10\,Myr, normalised to the binary fraction (0.42). 

The log-normal fit to the G-dwarf binaries by \citet{Raghavan10}, normalised to the observed binary fraction of 0.46 is shown by the solid red line in Fig.~\ref{cold_common_G_sepdist}. The initial separation distribution, normalised to the binary fraction at 0\,Myr (0.73) is 
shown by the open histogram, and the sepration distribution at 10\,Myr is shown by the hashed histogram, normalised to the binary fraction (0.55).

In Fig.~\ref{cold_common_A_sepdist} we show the log-normal fit to the observed visual A-star binaries \citep{deRosa12,DeRosa14}, which has a mean separation of $\bar{a} = 389$\,au, by the solid green line, normalised to the A-star visual binary fraction of 0.48 \citep{DeRosa14}.  
\citet{Duchene13b} discuss a bi-modal separation distribution for A-star binaries, which takes into account short-separation spectroscopic binaries in associations, which may make up a significant fraction of the field but have separations that the survey of \citet{DeRosa14} is not sensitive to.  
We show the bi-modal distribution in \citet{Duchene13b} by the purple dot-dashed line in Fig.~\ref{cold_common_A_sepdist}, which is normalised to a binary fraction of 0.70. The binaries at 0\,Myr in our simulations are shown by the open histogram (normalised to the initial binary fraction 
of 0.75) and the separation distribution of binaries remaining after 10\,Myr is shown by the hashed histogram (normalised to the binary fraction of 0.56).

In these histograms normalised to binary fractions, too many close ($<10$\,au) binaries are produced (and remain after dynamical evolution) for the brown dwarf and M-dwarf primary mass ranges (Figs.~\ref{cold_common_BD_sepdist}~and~\ref{cold_common_M_sepdist}). After dynamical evolution, 
too many G-dwarf binaries in the separation range 1 -- 200\,au are present (Fig.~\ref{cold_common_G_sepdist}), and the previously reported deficit of wide binaries in dense clusters is also apparent \citep{Parker11c}.  When we compare the evolution of the separation distribution of A-star binaries 
in our simulations, the common primordial separation distribution over-produces binaries in the range 1 -- 100\,au, and under-produces binaries in the range 100 -- 10$^5$\,au (see Fig.~\ref{cold_common_A_sepdist}). We also note that that a common primordial separation distribution is also inconsistent 
with the bi-modal separation distribution for A-stars presented in \citet{Duchene13b}. 

The histograms in Fig.~\ref{cold_common_sep_results_dist} show the evolution of both the binary fraction and the separation distribution. For completeness, we now examine only the evolution of the separation distributions for each primary mass range using cumulative distributions. The results are 
shown in Fig.~\ref{cold_common_sep_results_cum} where the fits to the data as observed in the Galactic Field are shown by the dashed lines in each panel. The alternative fit to the brown dwarf binary distribution 
by \citet{Basri06} is shown by the dot-dashed (magenta) line in Fig.~\ref{cold_common_BD_sepcum}, and the bimodal fit to the A-star data by \citet{Duchene13b} is shown by the dot-dashed (purple) line in Fig.~\ref{cold_common_A_sepcum}. 
In all panels, the separation distribution as measured at 0\,Myr in the simulations is shown by the dotted lines, and the distribution after 10\,Myr of dynamical evolution is shown by the solid lines. 

As indicated in the histograms in Fig.~\ref{cold_common_sep_results_dist}, far too many close binaries remain after dynamical evolution, which dominate the cumulative separation distributions. In the case of the G-dwarfs and A-stars, the distributions 
at 0\,Myr already sit to the left (closer separations) of the observed field distributions. The subsequent dynamical evolution then shifts these distributions to even closer separations on average. The only binaries which have a roughly similar distribution to the field are those with  
brown dwarf primaries, where the mean separation after 10\,Myr of evolution is similar to the mean in the observed separation distribution . However, the overall shape of the distribution in the simulations is wider than the more commonly adopted fit to the \citet{Burgasser07} data by \citet{Thies07} -- see also \citet{Duchene13b}; 
instead the separation distribution in the simulations is more similar to the (postulated) extended distribution from \citet{Basri06}.

In summary, the evolution of a common primordial binary population can account for a decreasing binary fraction as a function of decreasing primary mass, but not for the observed differences in both the mean, and shape, of the separation distribution between systems with different primary masses in the field.

\subsection{Field-like binary population}

Here we will focus on two sets of simulations with the same binary population, where we take the binary fractions and separation distributions observed in the Galactic field as initial conditions. We then evolve the star forming regions 
in two distinct ways; in one set of simulations the regions are subvirial (i.e.\,\,collapsing) and in the other they are supervirial (expanding) so that we can determine the fraction and properties of systems that form through capture during 
the dissolution of the regions \citep{Kouwenhoven10,Moeckel10}.

\subsubsection{Regions undergoing cool-collapse}
\label{results:fieldcollapse}

We first examine the evolution of an initially field-like binary population in a cool, collapsing star forming region. The results are summarised in Table~\ref{summary_cold_field}. 

\begin{figure}
\begin{center}
\rotatebox{270}{\includegraphics[scale=0.4]{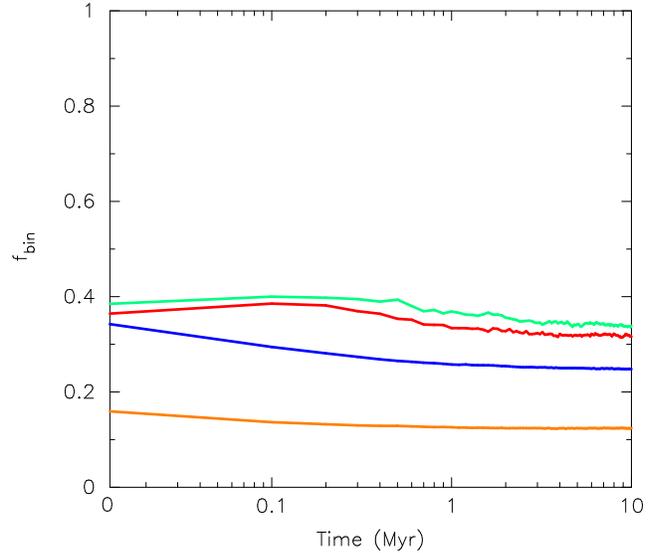}}
\end{center}
\caption[bf]{Evolution of the binary fraction for binaries with properties drawn from the field population distributions in simulated dense star forming regions undergoing cool-collapse. 
The first (top, green) line shows the evolution of the A-star binary fraction; the second (red) line shows the evolution of the G-dwarf binary fraction; the third (blue) line shows the 
evolution of the M-dwarf binary fraction; and the fourth (bottom, orange) line shows the evolution of the brown dwarf binary fraction. }
\label{cold_field_bin_frac}
\end{figure}

\begin{figure*}
  \begin{center}
\setlength{\subfigcapskip}{10pt}
\subfigure[Brown dwarf binaries, 10 Myr]{\label{cold_field_BD_sepdist}\rotatebox{270}{\includegraphics[scale=0.35]{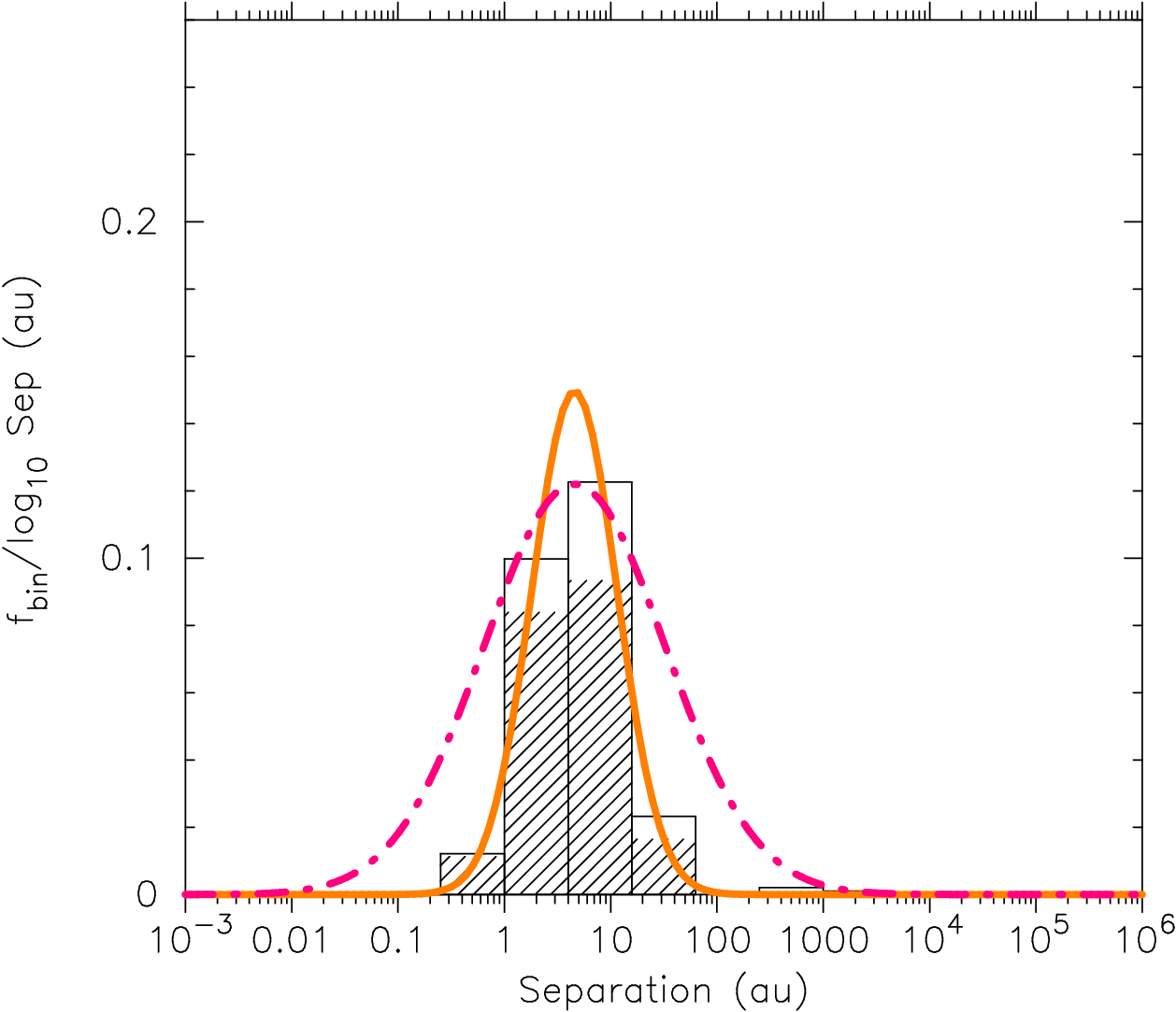}}}
\hspace*{0.8cm}
\subfigure[M-dwarf binaries, 10 Myr]{\label{cold_field_M_sepdist}\rotatebox{270}{\includegraphics[scale=0.35]{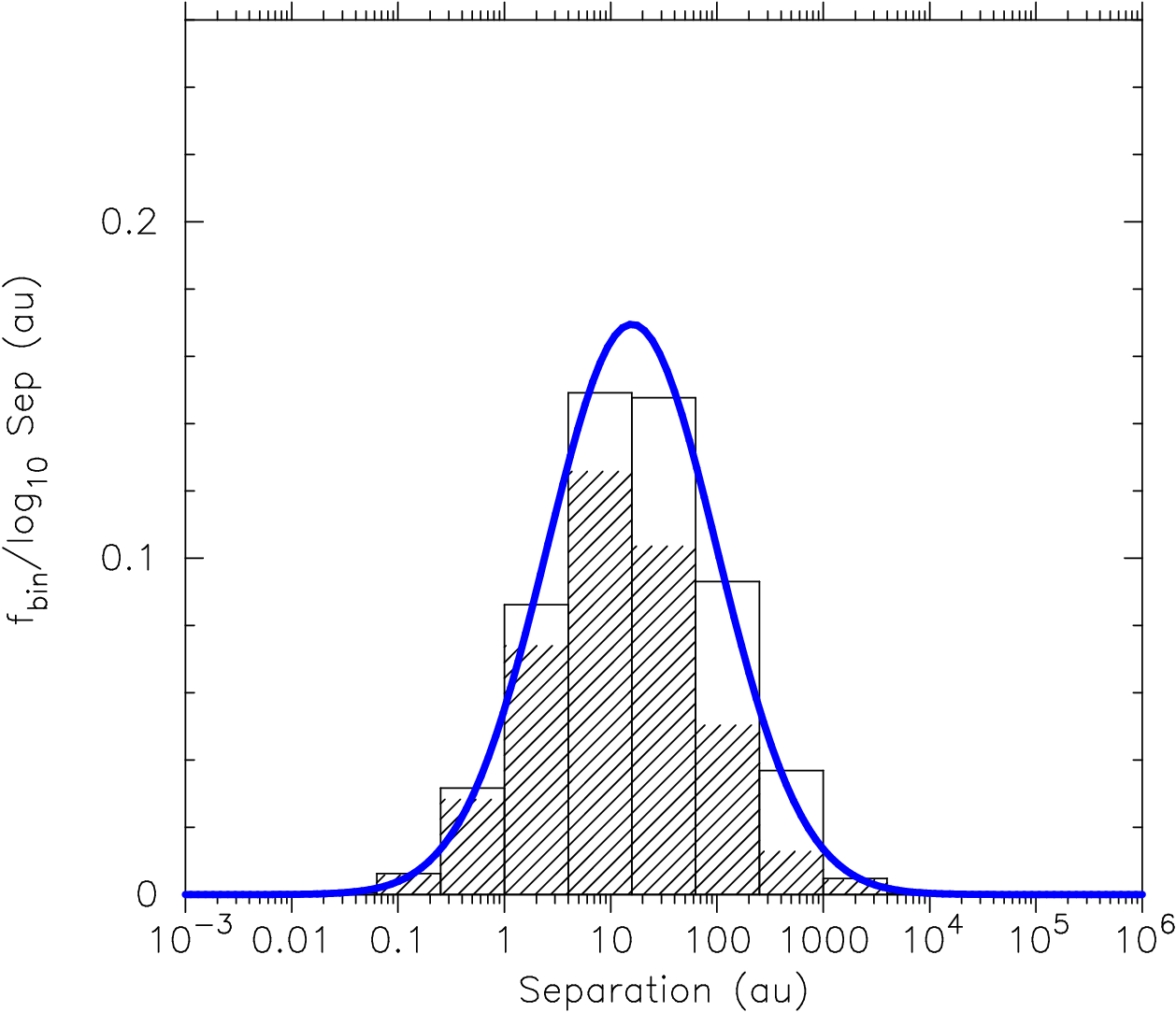}}}
\vspace*{0.25cm}
\subfigure[G-dwarf binaries, 10 Myr]{\label{cold_field_G_sepdist}\rotatebox{270}{\includegraphics[scale=0.35]{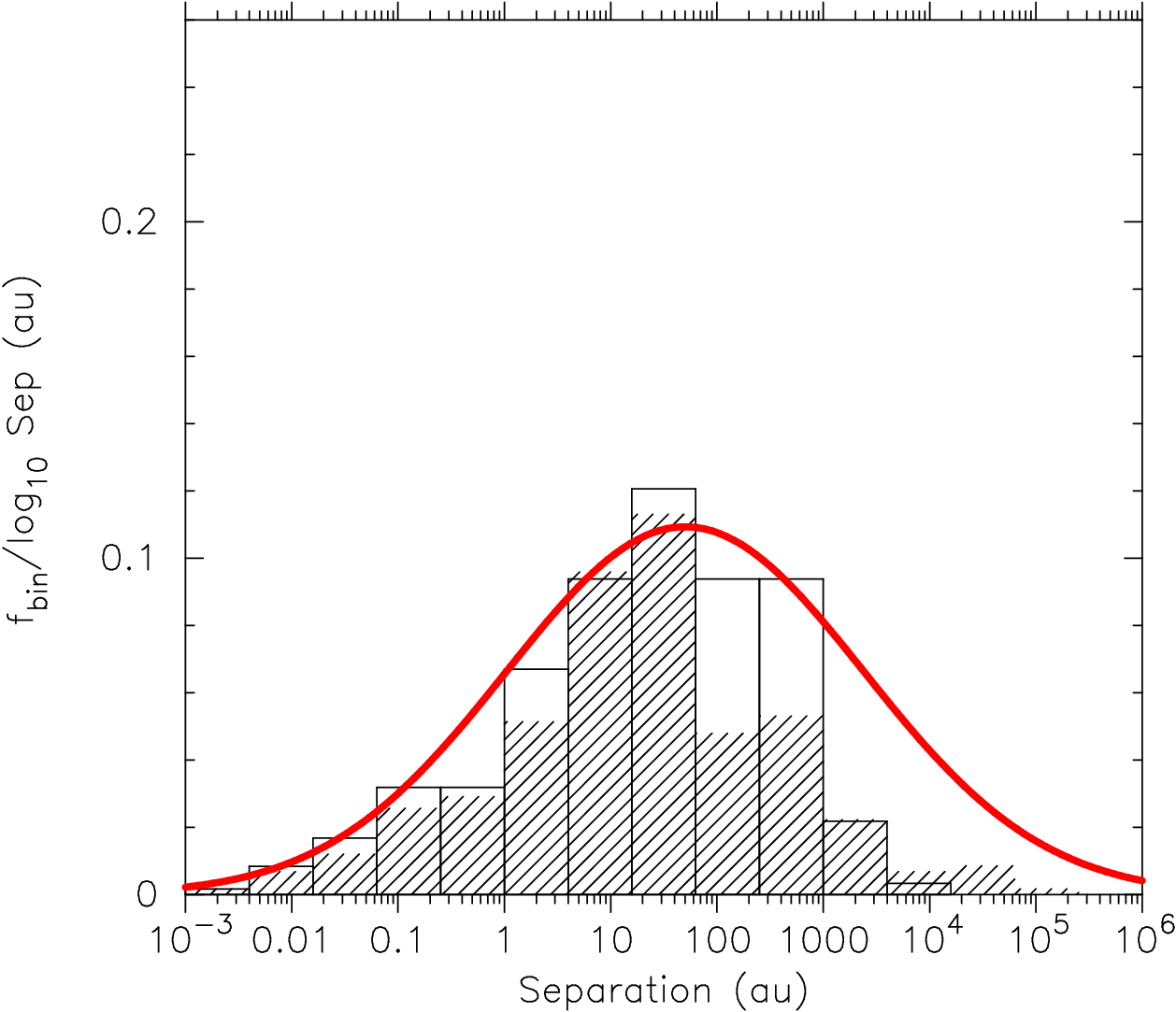}}}
\hspace*{0.8cm}
\subfigure[A-star binaries, 10 Myr]{\label{cold_field_A_sepdist}\rotatebox{270}{\includegraphics[scale=0.35]{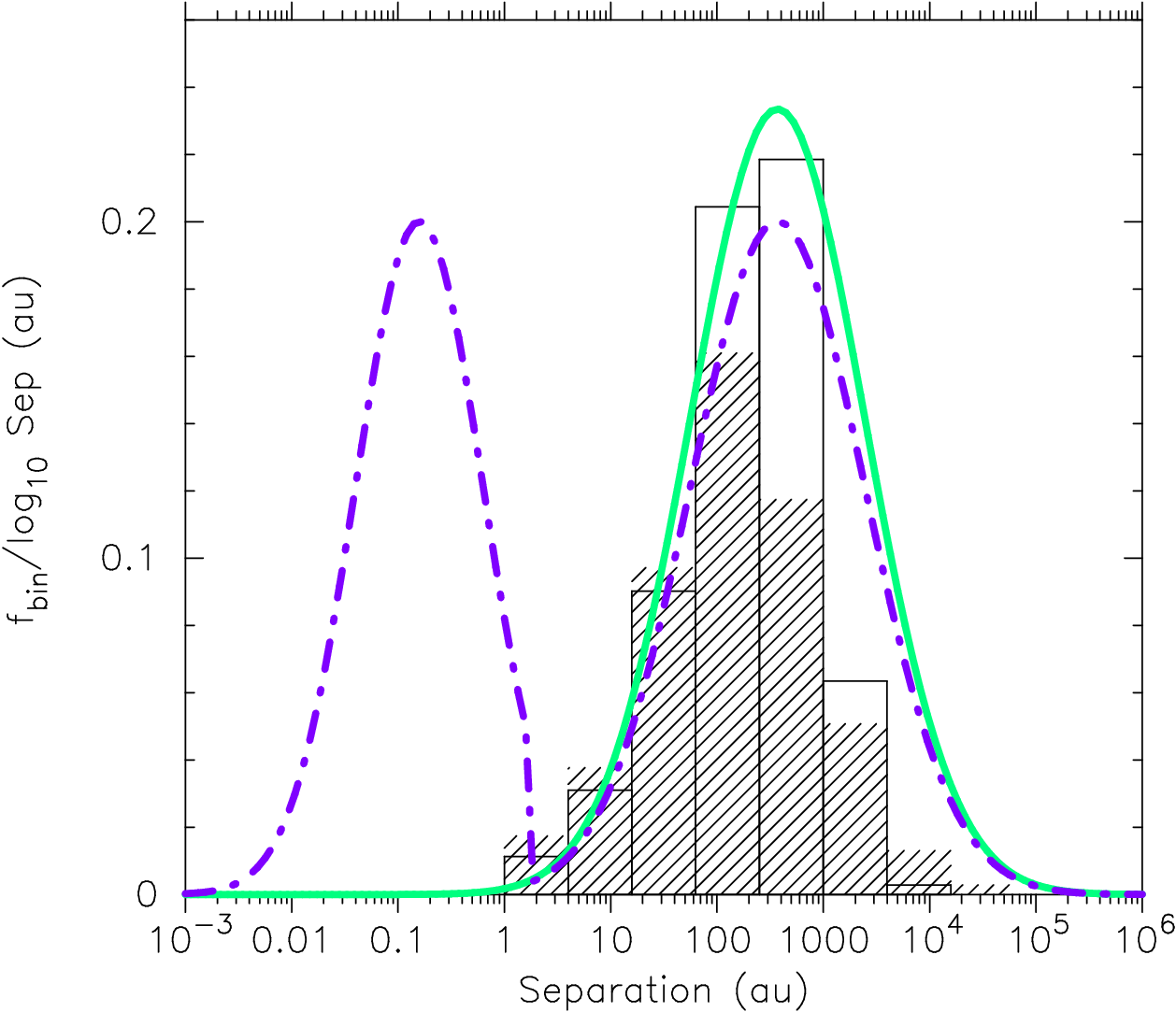}}}
  \end{center}
  \caption[bf]{Evolution of the separation distributions for binaries with properties drawn from the field population distributions in simulated dense star forming regions undergoing cool-collapse. In all panels, 
the open histogram shows the distribution at 0\,Myr (i.e. before dynamical evolution) and the hashed histogram shows the distribution after 10\,Myr. In panel 
(a) we show the evolution of the distribution for brown dwarf (BD) binaries; the log-normal approximation to the data from \citet{Thies07} is shown by the (solid) orange line (normalised to a binary fraction of 0.15), and the log-normal approximation 
to the data assuming `missing' systems \citep{Basri06} is shown by the (dot-dashed) magenta line (normalised to a binary fraction of 0.26). In panel (b) we show the evolution of the distribution for M-dwarf binaries; the log-normal 
approximation to the data by \citet{Janson12} is shown by the (solid) blue line (normalised to a binary fraction of 0.34). In panel (c) we show the evolution of the distribution for G-dwarf binaries; the log-normal approximation to the 
data by \citet{Raghavan10} is shown by the (solid) red line. In panel (d) we show the evolution of the distribution for A-star binaries; the log-normal approximation to the visual binary data by \citet{DeRosa14} is shown by the (solid) green line (normalised 
to a binary fraction of 0.48), and the fit to the bimodal distribution discussed in \citet{Duchene13b} is shown by the (dot-dashed) purple line.}
  \label{cold_field_sep_results_dist}
\end{figure*}

\begin{table*}
\caption[bf]{A summary of the results for the simulations of regions undergoing cool-collapse which contain binaries drawn from the field distributions (see Table~\ref{field_props} for details). From left to right, 
the columns are primary component mass-type, the input binary fraction, $f_{\rm bin}$ (init.), the actual binary fraction calculated before dynamical evolution, $f_{\rm bin}$ (0\,Myr), the binary fraction 
calculated after 10\,Myr of dynamical evolution, $f_{\rm bin}$ (10\,Myr), the median separation before dynamical evolution, $\tilde{a}$ (0\,Myr), and the median separation after 10\,Myr of 
dynamical evolution, $\tilde{a}$ (10\,Myr).}
\begin{center}
\begin{tabular}{|c|c|c|c|c|c|}
\hline 
Primary &  $f_{\rm bin}$ (init.) & $f_{\rm bin}$ (0\,Myr) & $f_{\rm bin}$ (10\,Myr) & $\tilde{a}$ (0\,Myr) & $\tilde{a}$ (10\,Myr) \\
\hline
A-star & 0.48  & 0.38 & 0.34 & 222\,au & 46\,au \\
\hline
G-dwarf & 0.46 & 0.36 & 0.32 & 25\,au & 18\,au \\
\hline
M-dwarf & 0.34 & 0.33 & 0.24 & 16\,au & 12\,au \\
\hline
Brown dwarf & 0.15 & 0.16 & 0.12 & 4.8\,au & 4.4\,au \\
\hline
\end{tabular}
\end{center}
\label{summary_cold_field}
\end{table*}

\begin{figure*}
  \begin{center}
\setlength{\subfigcapskip}{10pt}
\subfigure[Brown dwarf binaries, 10 Myr]{\label{cold_field_BD_sepcum}\rotatebox{270}{\includegraphics[scale=0.35]{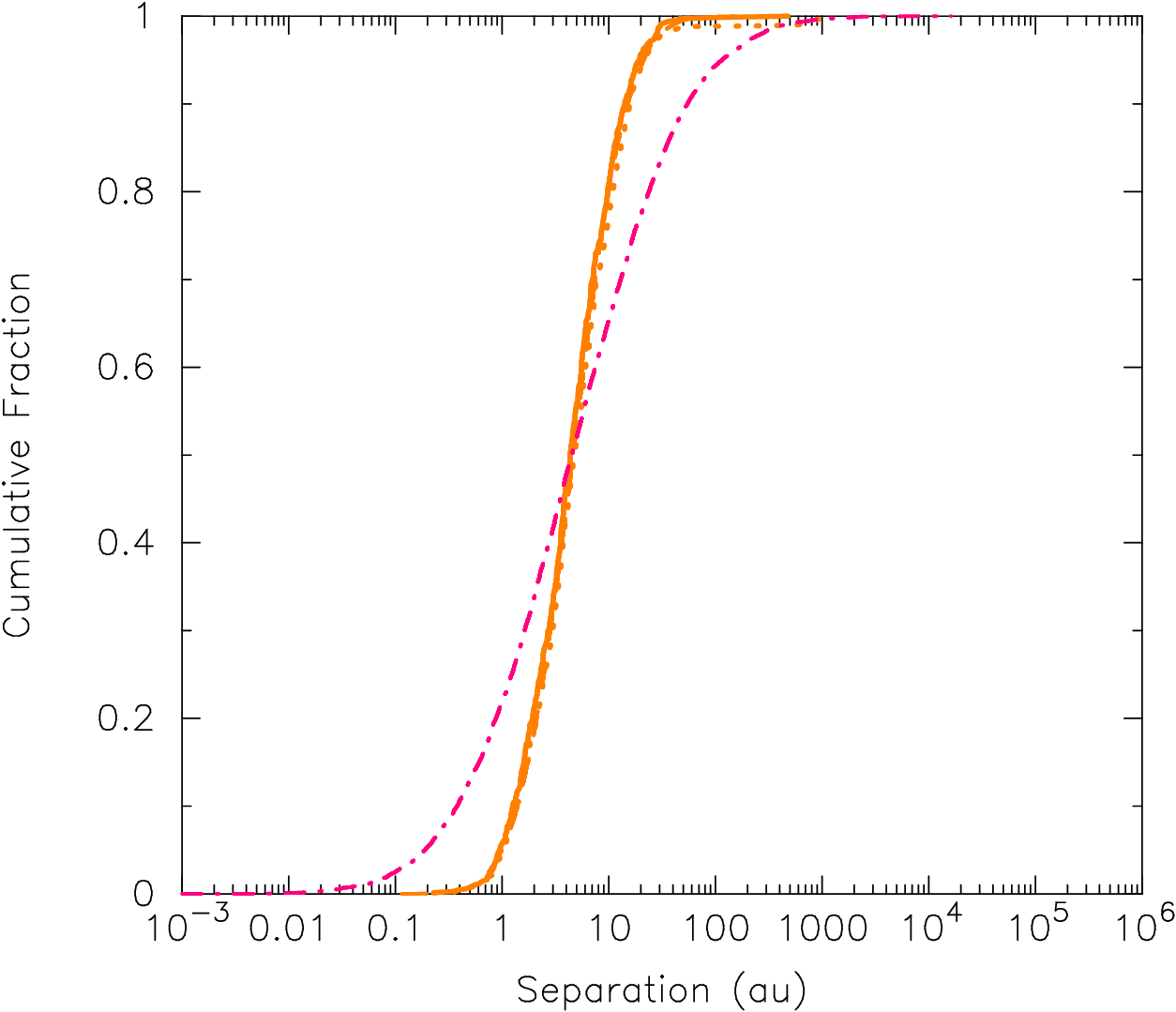}}}
\hspace*{0.8cm}
\subfigure[M-dwarf binaries, 10 Myr]{\label{cold_field_M_sepcum}\rotatebox{270}{\includegraphics[scale=0.35]{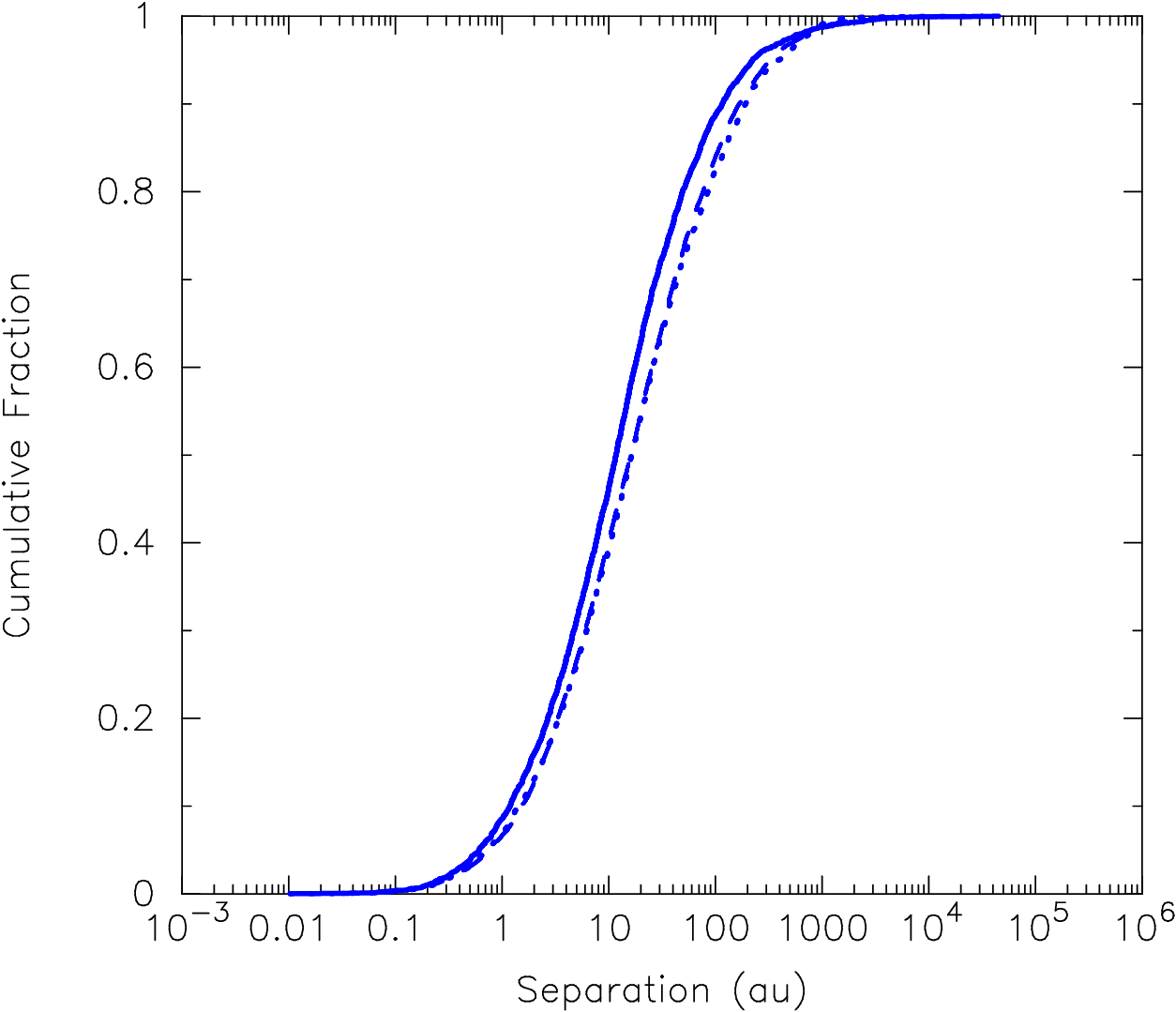}}}
\vspace*{0.25cm}
\subfigure[G-dwarf binaries, 10 Myr]{\label{cold_field_G_sepcum}\rotatebox{270}{\includegraphics[scale=0.35]{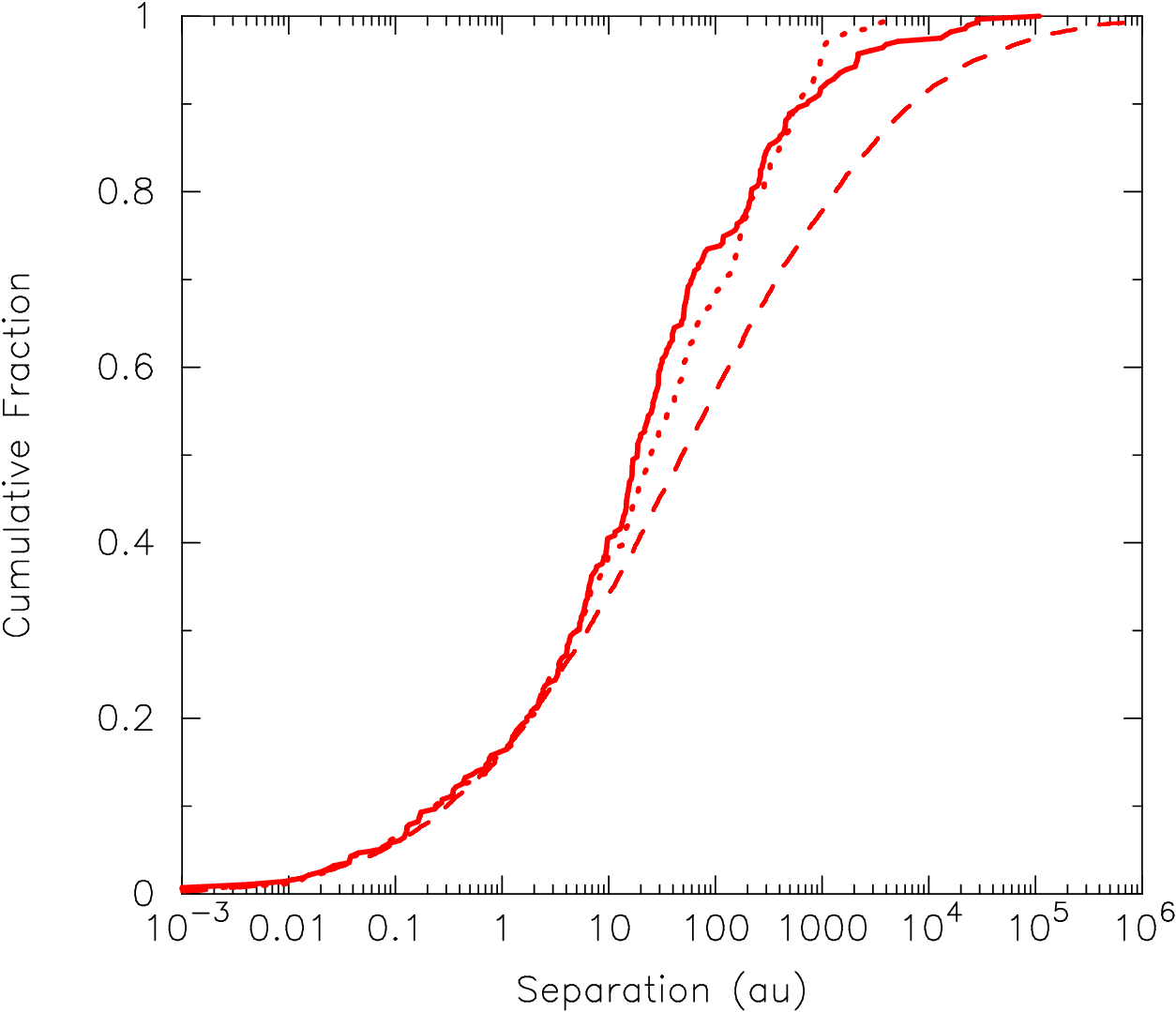}}}
\hspace*{0.8cm}
\subfigure[A-star binaries, 10 Myr]{\label{cold_field_A_sepcum}\rotatebox{270}{\includegraphics[scale=0.35]{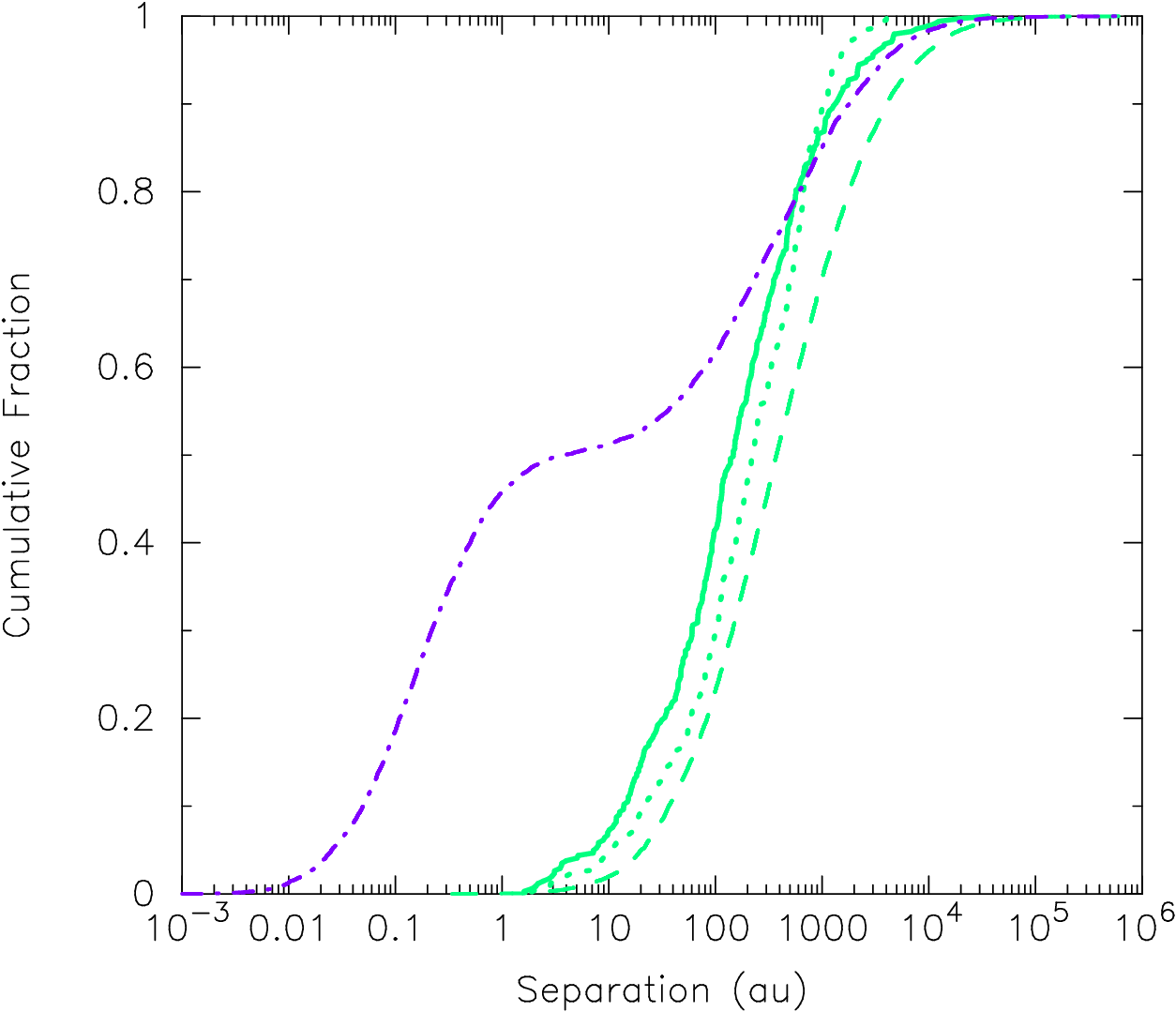}}}
  \end{center}
  \caption[bf]{Evolution of the cumulative separation distributions for binaries with properties drawn from the field population distributions in simulated dense star forming regions undergoing cool-collapse. In all panels, 
the dotted line shows the distribution at 0\,Myr (i.e. before dynamical evolution) and the solid line shows the distribution after 10\,Myr. In all panels the dashed lines show the respective cumulative distributions of the log-normal fits to the data for each primary mass range observed in the Galactic field 
(detailed in Table~\ref{field_props}). In panel 
(a) the cumulative distribution proposed by \citep{Basri06} is shown by the dot-dashed magenta line. In panel (d) the bimodal distribution discussed in \citet{Duchene13b} is shown by the dot-dashed purple line.}
  \label{cold_field_sep_results_cum}
\end{figure*}

Fig.~\ref{cold_field_bin_frac} shows the evolution of the binary fraction over 10\,Myr for our four chosen primary mass ranges. As in the case of the common primordial binary population, the calculated binary fractions of the A-stars (the top green line) and G-dwarfs (upper middle red line) before dynamical 
evolution (0\,Myr) are lower than the input values, because the widest binaries are not physically bound in the high density star-forming regions. The M-dwarf and brown dwarf binaries are generally so close that they are all bound before dynamical evolution. In fact, in this suite of simulations, 
the calculated brown dwarf binary fraction is slightly higher than the input fraction, likely due to wide pairs of brown dwarfs being classified as binaries due to the correlated velocities in the fractals [this is evident in the bin between 500 -- 1000\,au in the separation distribution (Fig.~\ref{cold_field_BD_sepdist}), 
which lies outside the range of the log-normal distribution used to generate the input separations].

During the subsequent dynamical evolution, the binary fractions are reduced from their initial values. The greatest change occurs for M-dwarf primaries, where the binary fraction is lowered from 0.33 (0\,Myr) to 0.24 (10\,Myr). This is slightly less 
than the change reported in \citet{Parker11c} for similar initial conditions for star forming regions. However, in that paper the input binary distribution contained a far higher proportion of wide M-dwarf systems 
because their separations were drawn from the \citet{Fischer92} fit to the M-dwarf separation distribution.

The reduction of  the binary fractions of A-star, G-dwarf and brown dwarf primaries are minimal (0.07, 0.06 and 0.04, respectively).

The separation distributions, normalised to binary fraction are shown in Fig.~\ref{cold_field_sep_results_dist}. In all cases, the open histograms show the separation distribution before dynamical evolution (0\,Myr) and after 10\,Myr (the hashed histogram). The log-normal fits to the observed 
data are shown by the solid lines in each panel, and the alternative postulated fits to the brown dwarf binaries and the A-star binaries are shown by the dot-dashed lines in Figs.~\ref{cold_field_BD_sepdist}~and~\ref{cold_field_A_sepdist}, respectively. As detailed in Section~\ref{results:common}, 
these histograms combine the evolution of the binary fraction \emph{and} separation distribution. The brown dwarf binary fraction decreases, 
but the shape of the distribution is still consistent with that observed in the field (Fig.~\ref{cold_field_BD_sepdist}). The M-dwarf, G-dwarf and A-star distributions show that too few wide ($>$100\,au) binaries remain after dynamical evolution, although in the case of the G-dwarfs and A-stars, a 
substantial fraction of systems from the input distribution are not physically bound before dynamical evolution takes place.

\begin{table*}
\caption[bf]{A summary of the results for the simulations of regions undergoing warm expansion which contain binaries drawn from the field distributions (see Table~\ref{field_props} for details). From left to right, 
the columns are primary component mass-type, the input binary fraction, $f_{\rm bin}$ (init.), the actual binary fraction calculated before dynamical evolution, $f_{\rm bin}$ (0\,Myr), the binary fraction 
calculated after 10\,Myr of dynamical evolution, $f_{\rm bin}$ (10\,Myr), the median separation before dynamical evolution, $\tilde{a}$ (0\,Myr), and the median separation after 10\,Myr of 
dynamical evolution, $\tilde{a}$ (10\,Myr).}
\begin{center}
\begin{tabular}{|c|c|c|c|c|c|}
\hline 
Primary &  $f_{\rm bin}$ (init.) & $f_{\rm bin}$ (0\,Myr) & $f_{\rm bin}$ (10\,Myr) & $\tilde{a}$ (0\,Myr) & $\tilde{a}$ (10\,Myr) \\
\hline
A-star & 0.48  & 0.36 & 0.48 & 234\,au & 239\,au \\
\hline
G-dwarf & 0.46 & 0.38 & 0.40 & 23\,au & 39\,au \\
\hline
M-dwarf & 0.34 & 0.33 & 0.27 & 16\,au & 13\,au \\
\hline
Brown dwarf & 0.15 & 0.15 & 0.12 & 4.8\,au & 4.6\,au \\
\hline
\end{tabular}
\end{center}
\label{summary_warm_field}
\end{table*}

When we examine the the cumulative separation distributions (Fig.~\ref{cold_field_sep_results_cum}), we see that the shape of the brown dwarf separation distribution does not deviate from the input (field) distribution (Fig.~\ref{cold_field_BD_sepcum}), and the alteration of the M-dwarf separation distribution 
Fig.~\ref{cold_field_M_sepcum}) is not as drastic as suggested when the data are binned and normalised to the binary fraction, as in the histogram in Fig.~\ref{cold_field_M_sepdist}. 

The G-dwarf separation distribution contains slightly too few wide binaries, even before dynamical evolution, which is apparent when comparing the simulation at 0\,Myr (dotted line) to the observed distribution in the field (the dashed line) in Fig.~\ref{cold_field_G_sepcum}. The subsequent dynamical evolution (shown by the solid line) 
shows two effects; the destruction of intermediate-wide binaries (with separations in the range 10 -- 1000\,au), but also the formation of very wide binaries ($>$1000\,au) as the cluster expands after collapse. A similar effect is also found for the A-stars (the solid line in Fig.~\ref{cold_field_A_sepcum}), and as with the G-dwarfs, 
the observed field distribution (the dashed line) lies to the right (wider separations) than both the 0\,Myr and 10\,Myr distributions in the simulations.

\subsubsection{Regions undergoing warm expansion}
\label{results:warmexpand}

Now we examine the evolution of a field-like binary population in a warm, expanding star-forming region. The dynamical evolution of these regions (without primordial binaries) was studied in detail by \citet{Parker14b}, and we refer the interested reader to that work for 
further details. The results are summarised in Table~\ref{summary_warm_field}.

\begin{figure}
\begin{center}
\rotatebox{270}{\includegraphics[scale=0.4]{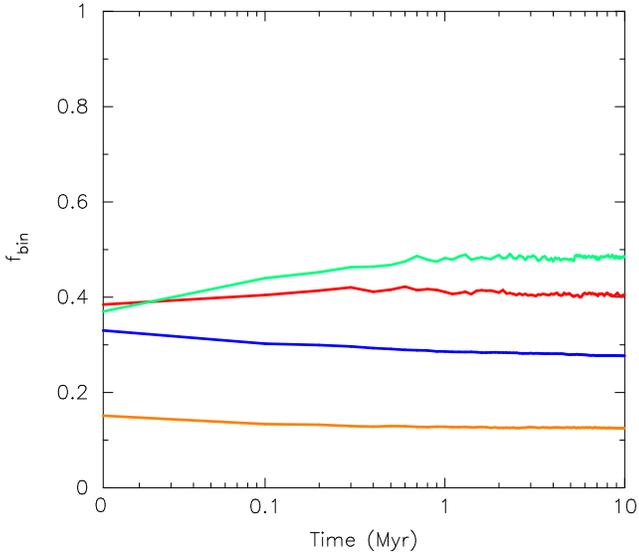}}
\end{center}
\caption[bf]{Evolution of the binary fraction for binaries with properties drawn from the field population distributions in simulated dense star forming regions undergoing warm expansion. 
The first (top, green) line shows the evolution of the A-star binary fraction; the second (red) line shows the evolution of the G-dwarf binary fraction; the third (blue) line shows the 
evolution of the M-dwarf binary fraction; and the fourth (bottom, orange) line shows the evolution of the brown dwarf binary fraction.}
\label{warm_field_bin_frac}
\end{figure}

In Fig.~\ref{warm_field_bin_frac} we show the evolution of the binary fraction over 10\,Myr for these expanding regions. The most striking feature of this plot is that, whilst the initially dense substructure processes some of the binaries, during the 
expansion of these regions the more massive stars form binary systems through capture \citep{Kouwenhoven10,Moeckel10}. This is seen in the 
evolution of the binary fractions; the A-star fraction rises from 0.36 to 0.48 and the G-dwarf fraction also increases slightly. 

\begin{figure*}
  \begin{center}
\setlength{\subfigcapskip}{10pt}
\subfigure[Brown dwarf binaries, 10 Myr]{\label{warm_field_BD_sepdist}\rotatebox{270}{\includegraphics[scale=0.35]{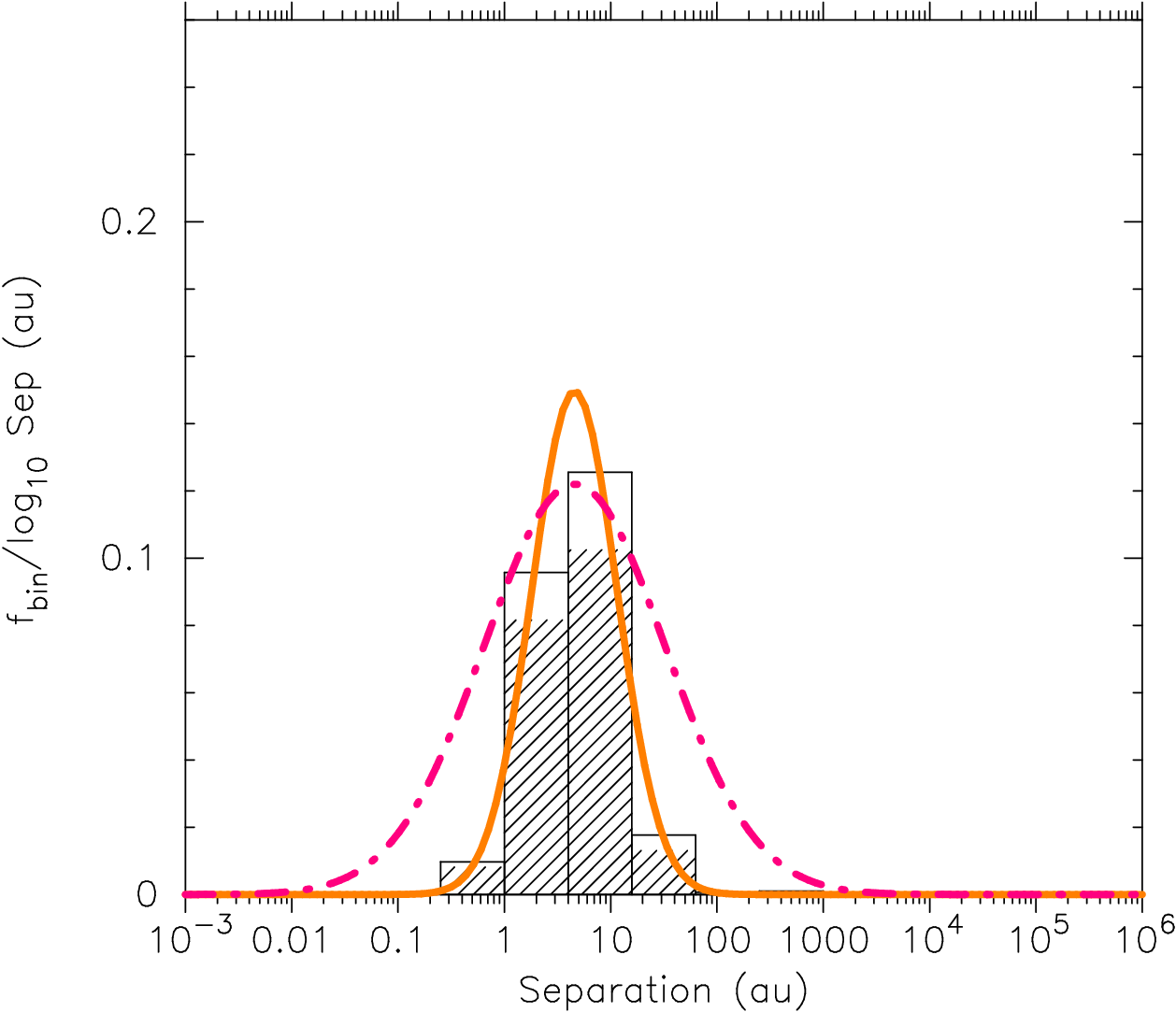}}}
\hspace*{0.8cm}
\subfigure[M-dwarf binaries, 10 Myr]{\label{warm_field_G_sepdist}\rotatebox{270}{\includegraphics[scale=0.35]{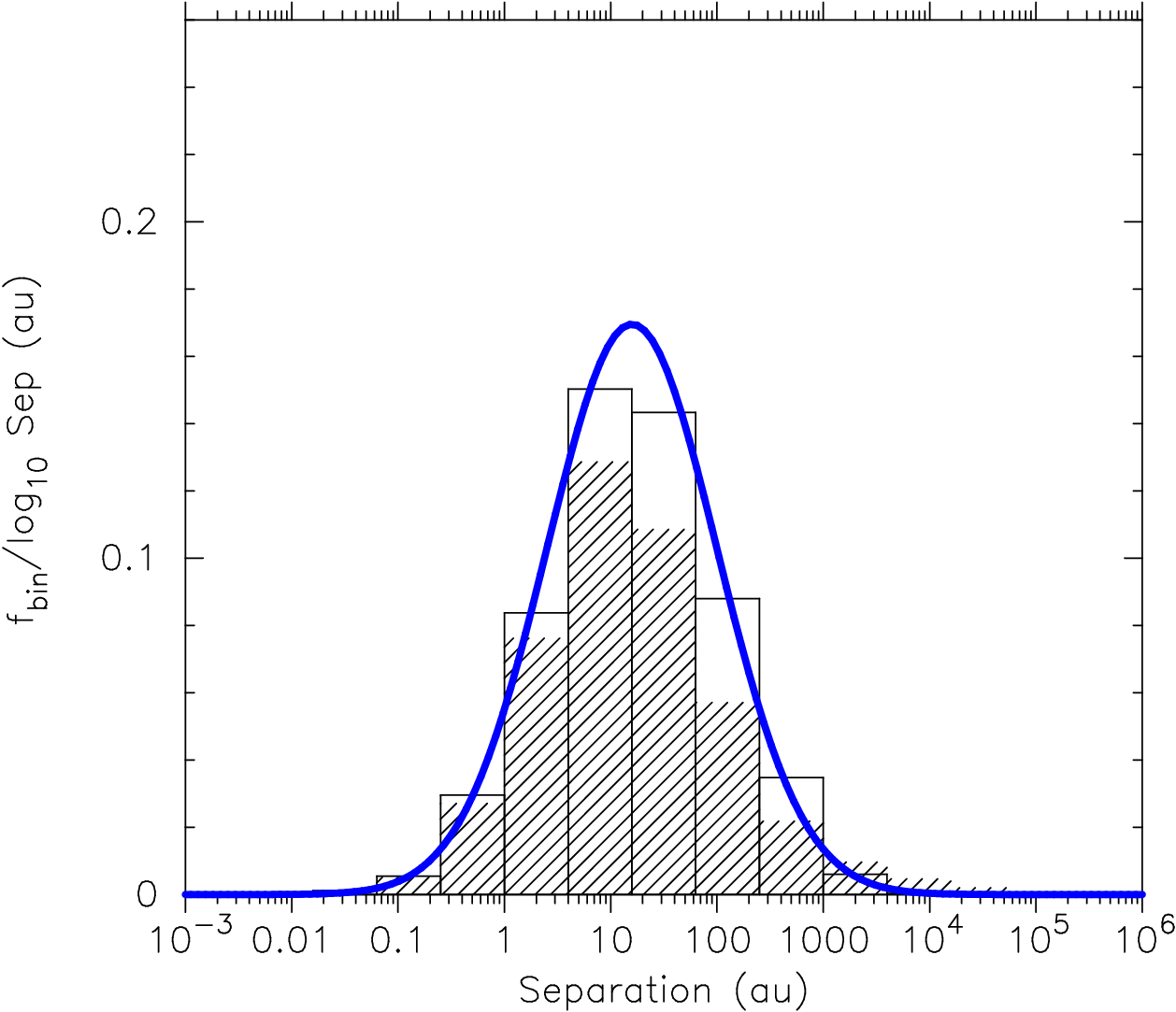}}}
\vspace*{0.25cm}
\subfigure[G-dwarf binaries, 10 Myr]{\label{warm_field_M_sepdist}\rotatebox{270}{\includegraphics[scale=0.35]{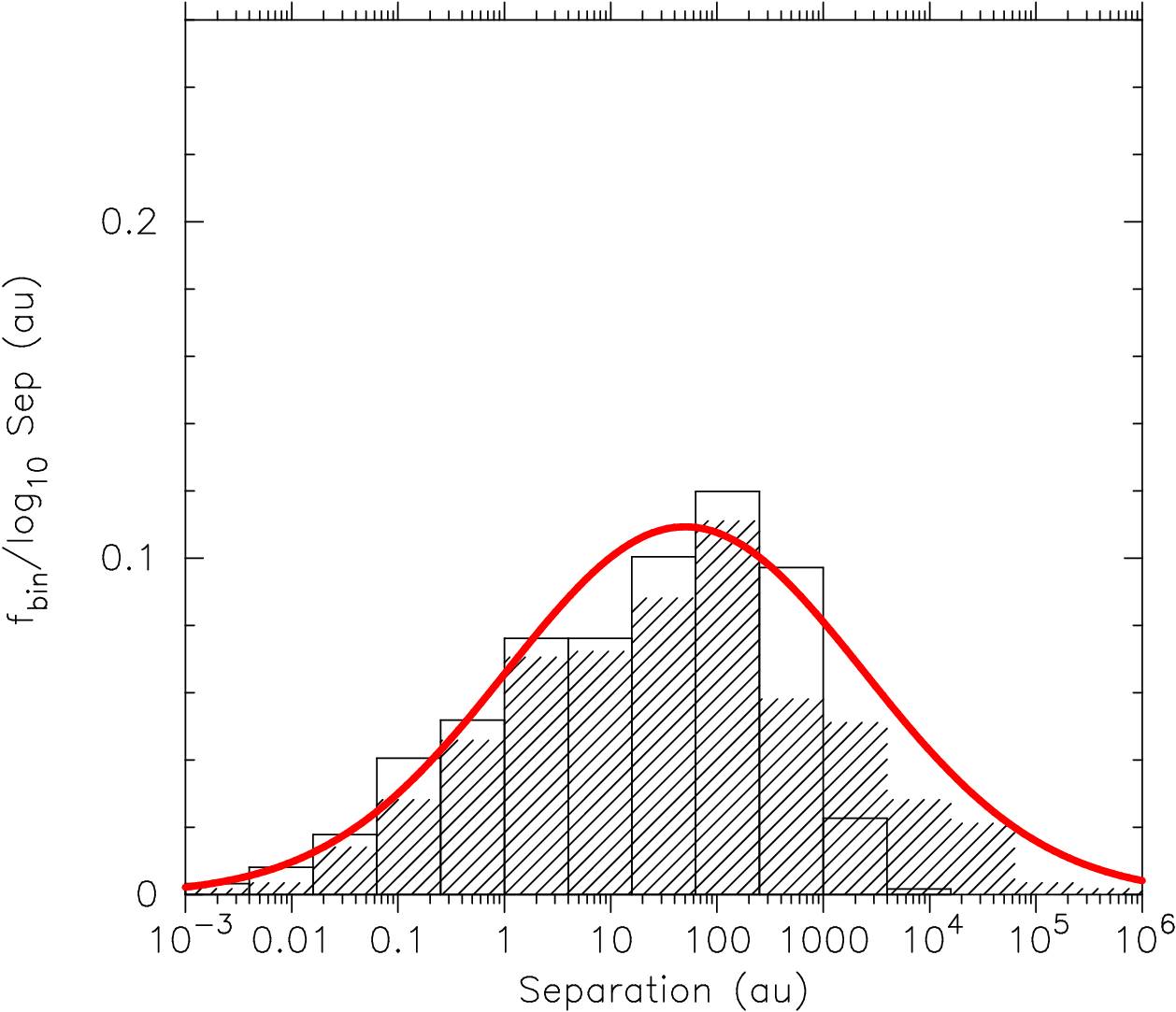}}}
\hspace*{0.8cm}
\subfigure[A-star binaries, 10 Myr]{\label{warm_field_A_sepdist}\rotatebox{270}{\includegraphics[scale=0.35]{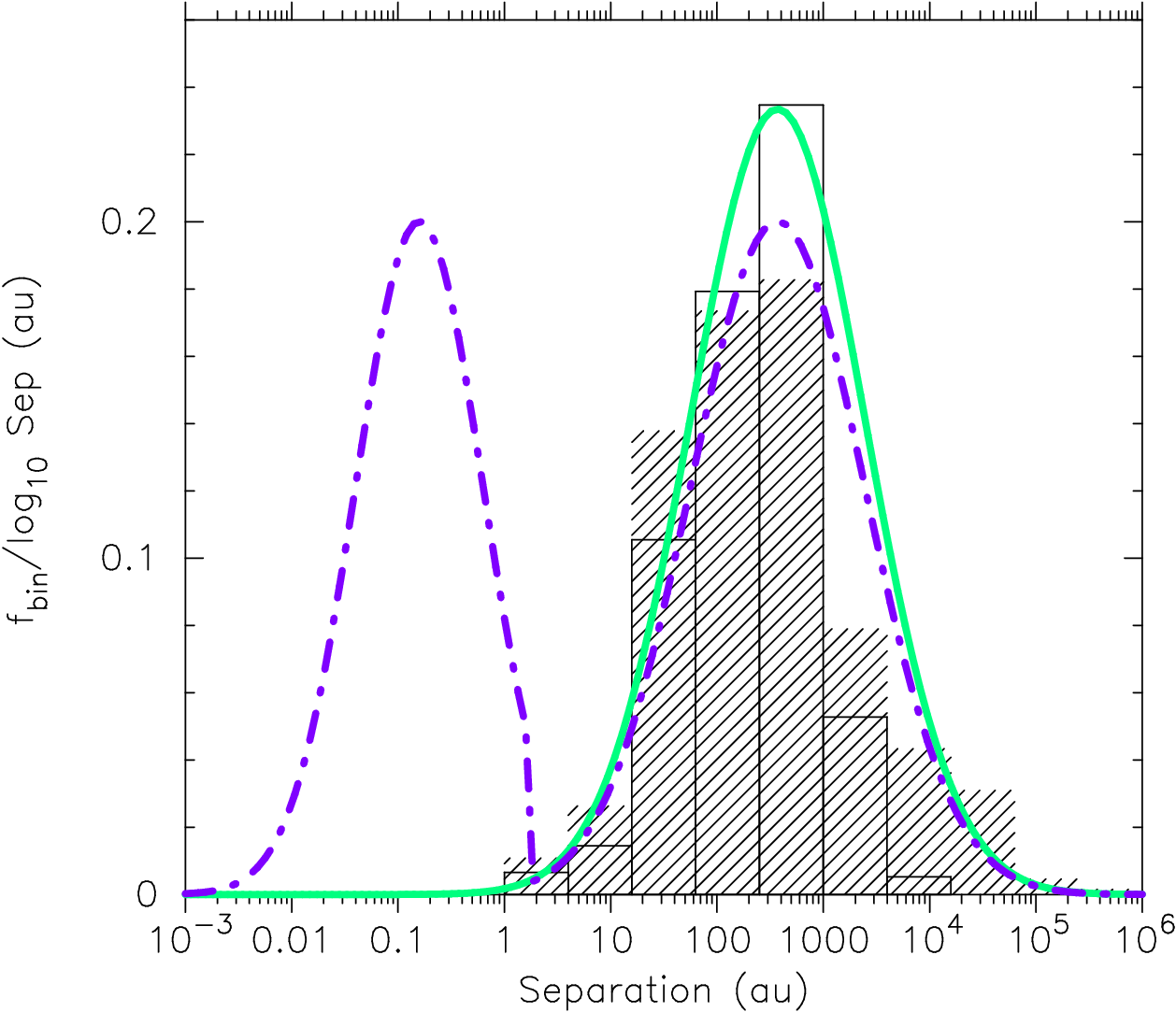}}}
  \end{center}
  \caption[bf]{Evolution of the separation distributions for binaries with properties drawn from the field population distributions in simulated dense star forming regions undergoing warm expansion. In all panels, 
the open histogram shows the distribution at 0\,Myr (i.e. before dynamical evolution) and the hashed histogram shows the distribution after 10\,Myr. In panel 
(a) we show the evolution of the distribution for brown dwarf (BD) binaries; the log-normal approximation to the data from \citet{Thies07} is shown by the (solid) orange line (normalised to a binary fraction of 0.15), and the log-normal approximation 
to the data assuming `missing' systems \citep{Basri06} is shown by the (dot-dashed) magenta line (normalised to a binary fraction of 0.26). In panel (b) we show the evolution of the distribution for M-dwarf binaries; the log-normal 
approximation to the data by \citet{Janson12} is shown by the (solid) blue line (normalised to a binary fraction of 0.34). In panel (c) we show the evolution of the distribution for G-dwarf binaries; the log-normal approximation to the 
data by \citet{Raghavan10} is shown by the (solid) red line. In panel (d) we show the evolution of the distribution for A-star binaries; the log-normal approximation to the visual binary data by \citet{DeRosa14} is shown by the (solid) green line (normalised 
to a binary fraction of 0.48), and the fit to the bimodal distribution discussed in \citet{Duchene13b} is shown by the (dot-dashed) purple line.}
  \label{warm_field_sep_results_dist}
\end{figure*}

The histograms of the evolution of the separation distributions in Fig.~\ref{warm_field_sep_results_dist} clearly show the destruction of some binaries (in all panels), but also the formation of wider systems in panels (c) and (d). This formation of these wider binaries 
is highly mass dependent; virtually no brown dwarf and M-dwarf binaries form during the regions' expansion, whereas significant numbers of G-dwarf and A-star binaries form. We interpret this as being due to the more massive G- and A-stars having a higher collisional 
crosss section, and are therefore more likely to form binaries via capture as the star forming region dissolves into the field. One potential caveat here is that the input separation distribution in the simulations contains many wide G-dwarf and A-star binaries, so these 
binaries which `form' during the dissolution may just be quasi-primordial binaries that become bound when the regions attain lower density. We exclude this possibility for two reasons. Firstly, we can tag primordial systems in our simulations, and the systems which form 
binaries do not always do so with their birth partner. In these simulations, after 10\,Myr and over all separations, 100\,per cent of brown dwarf primaries are with their birth partner, and the fraction decreases to 93\,per cent for M-dwarfs, 72\,per cent for G-dwarf and 58\,per cent for A-stars (i.e.\,\,42\,per cent of A-star binaries are not birth binaries). If we restrict our separation range of interest to $>$1000\,au, then no brown dwarf binaries were born in this separation range, only 21\,per cent of M-dwarf primaries are with their birth partner, rising to 28\,per cent for G-dwarfs and 27\,per cent for A-stars (the reason for this trend is that fewer M-dwarfs can form 1000\,au binaries than can G-dwarfs or A-stars due to the respective shapes of the separation distributions). Secondly, we ran a further suite of simulations with no primordial binaries \citep[similar to those in][]{Kouwenhoven10} and also found an increase in the number of wide binaries as a function of increasing primary mass 
which form over 10\,Myr.

\begin{figure*}
  \begin{center}
\setlength{\subfigcapskip}{10pt}
\subfigure[Brown dwarf binaries, 10 Myr]{\label{warm_field_BD_sepcum}\rotatebox{270}{\includegraphics[scale=0.35]{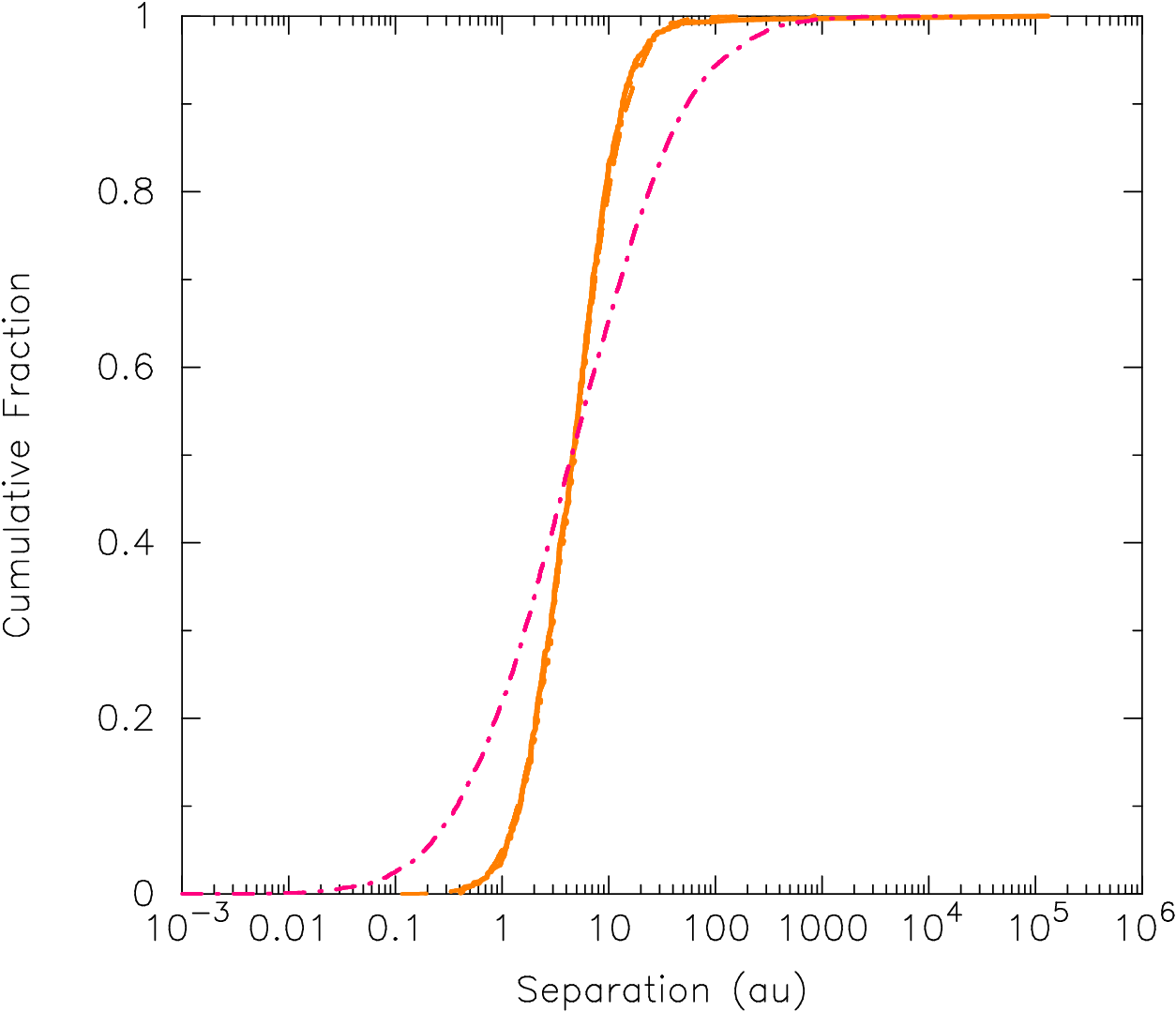}}}
\hspace*{0.8cm}
\subfigure[M-dwarf binaries, 10 Myr]{\label{warm_field_M_sepcum}\rotatebox{270}{\includegraphics[scale=0.35]{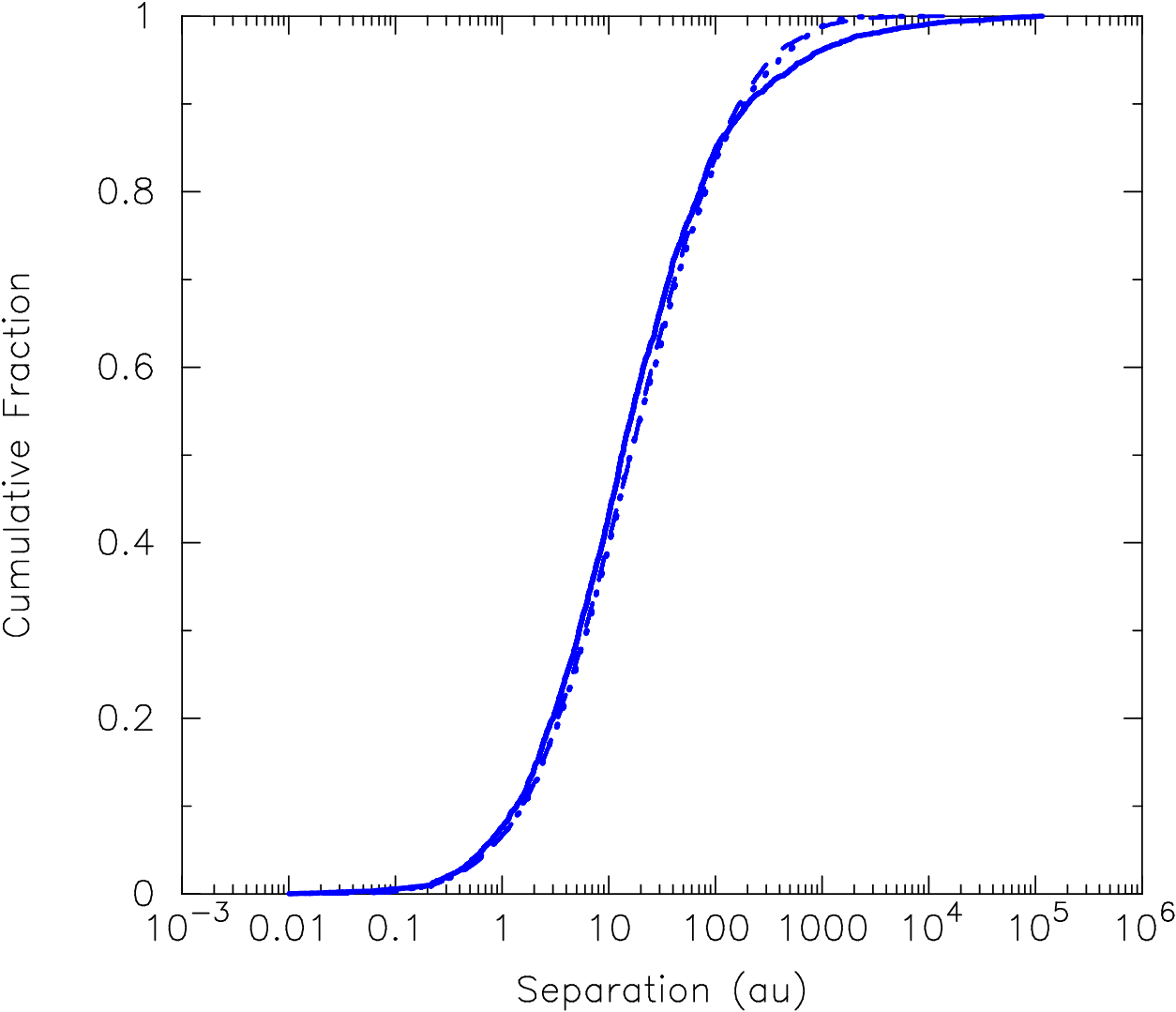}}}
\vspace*{0.25cm}
\subfigure[G-dwarf binaries, 10 Myr]{\label{warm_field_G_sepcum}\rotatebox{270}{\includegraphics[scale=0.35]{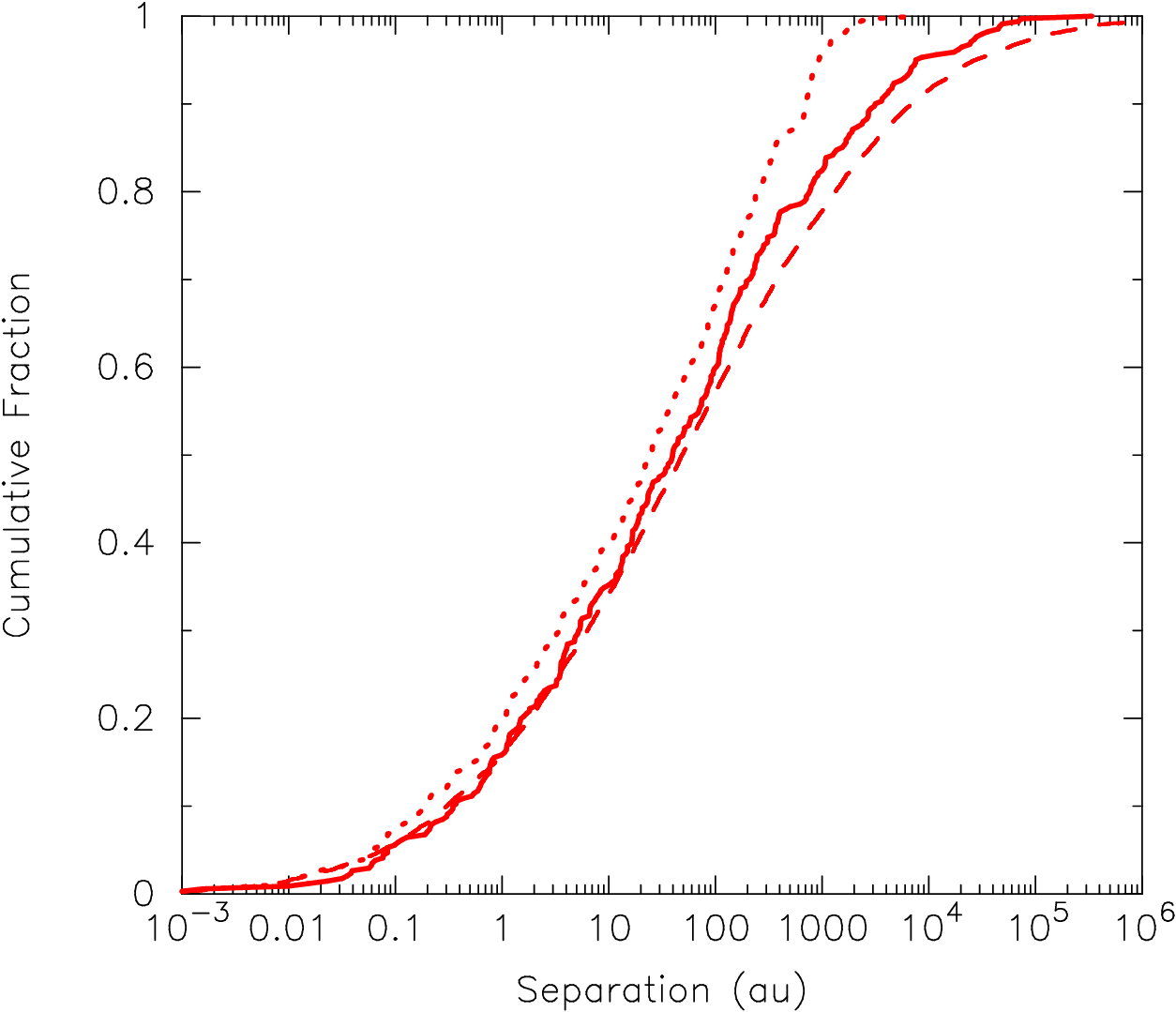}}}
\hspace*{0.8cm}
\subfigure[A-star binaries, 10 Myr]{\label{warm_field_A_sepcum}\rotatebox{270}{\includegraphics[scale=0.35]{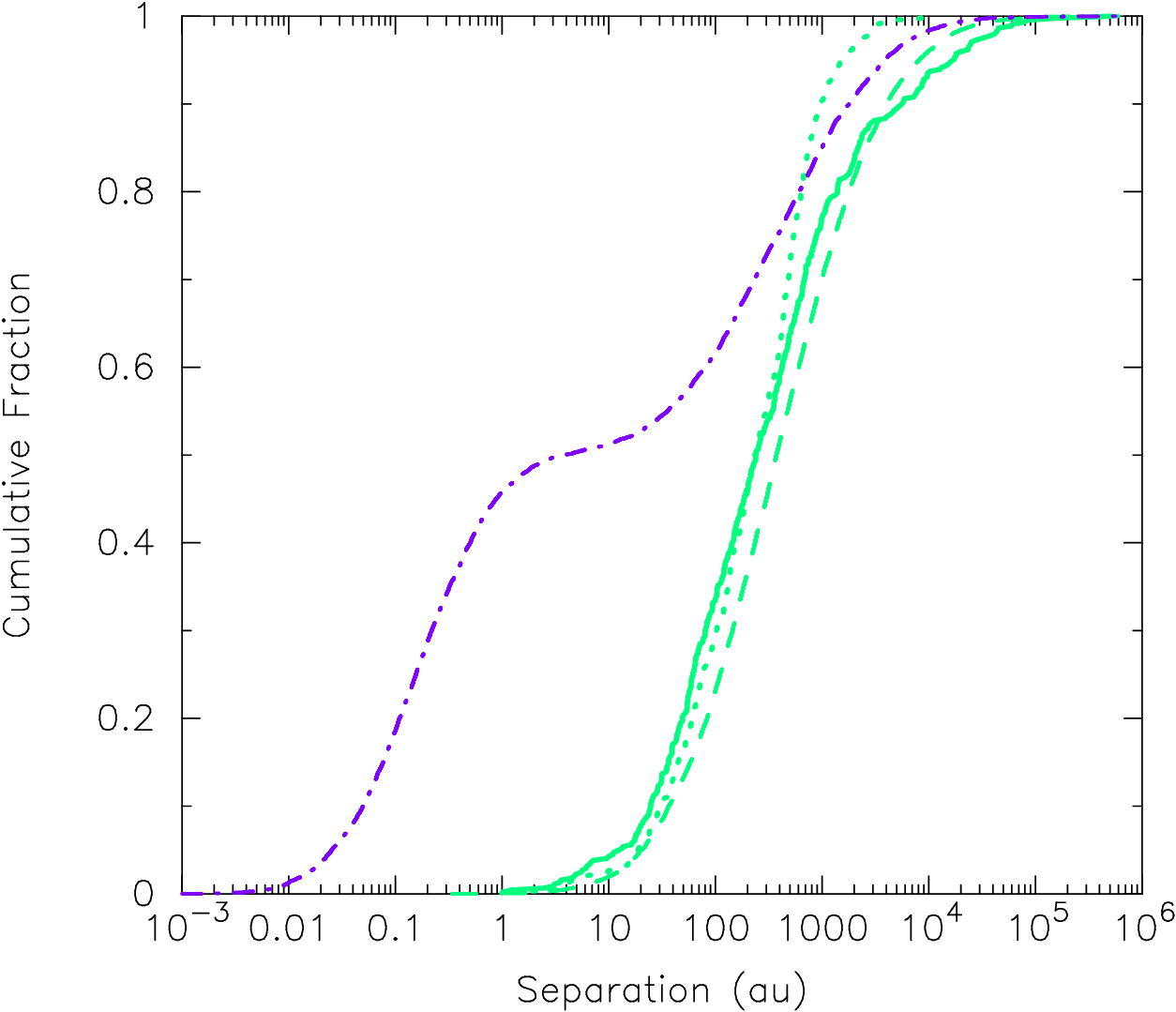}}}
  \end{center}
  \caption[bf]{Evolution of the cumulative separation distributions for binaries with properties drawn from the field population distributions in simulated dense star forming regions undergoing warm expansion. In all panels, 
the dotted line shows the distribution at 0\,Myr (i.e. before dynamical evolution) and the solid line shows the distribution after 10\,Myr. In all panels the dashed lines show the respective cumulative distributions of the log-normal fits to the data for each primary mass range observed in the Galactic field 
(detailed in Table~\ref{field_props}). In panel 
(a) the cumulative distribution proposed by \citep{Basri06} is shown by the dot-dashed magenta line. In panel (d) the bimodal distribution discussed in \citet{Duchene13b} is shown by the dot-dashed purple line.}
  \label{warm_field_sep_results_cum}
\end{figure*}

This is confirmed when we examine the cumulative separation distributions (Fig.~\ref{warm_field_sep_results_cum}), which show almost no change to the shape of the separation distributions for brown dwarf and M-dwarf binaries, but a substantial 
number of wide G-dwarf and A-star binaries form during the regions' dissolution.

Note that we have not considered the formation of hierarchical systems (triples, quadruples, etc) in our analysis; a high fraction of wide binaries in the field may actually be such systems and thus counted as `binary systems'. Any such 
systems in our simulations would merely reinforce the results described above.

\section{Discussion}
\label{discuss}

The $N$-body simulations presented in Section~\ref{results} show that the field binary population cannot be explained by the dynamical evolution of one single, common primordial population (i.e.\,\,binary fraction and separation distribution). Whilst dynamical evolution can explain the decreasing binary fraction with primary mass, 
it cannot account for the decreasing peak in mean separation with decreasing primary mass. The main problem is that in order to sculpt the initial separation distribution to match the observed deficit of intermediate/wide ($>$100\,au) low-mass (brown dwarf and M-dwarf) binary systems, the SF regions need to be so dense that 
the resultant dynamical evolution results in too few G-dwarf and A-star binaries with intermediate/wide separations remaining. If some SF regions are initially supervirial, correlated velocities on local scales \citep{Larson82} can result in the formation of wide G-dwarf and A-star binaries \citep{Kouwenhoven10,Moeckel10}, but then 
do not destroy enough hard/intermediate M-dwarf and brown dwarf binaries.

The above findings appear to contradict the earlier work by \citet{Kroupa95a,Kroupa95b,Kroupa08,Marks12}, who find that a common primordial binary population can explain the field binary population. This earlier work compared the results of $N$-body simulations to the G-dwarf and M-dwarf binary distributions 
observed in the field by \citet{Duquennoy91} and \citet{Fischer92}. Whilst the updated G-dwarf separation distribution and fraction presented by \citet{Raghavan10} does not differ greatly from \citet{Duquennoy91}, the improved M-dwarf surveys by \citet{Bergfors10} and \Citet{Janson12} suggest that both the binary fraction, and 
separation distribution lie inbetween the fractions/separation distributions of brown dwarf and G-dwarf binaries. Interestingly, the data presented by \citet{Fischer92} did suggest this, but only the recent surveys by \citet{Bergfors10} and \citet{Janson12} were able to demonstrate a clear difference from the G-dwarf distribution. 
 When compared to these updated observations, the hypothesised common primordial binary population from \citet{Kroupa95a,Kroupa11} is incompatible with the observations (regardless of primary mass range). 

If the primordial binary population used as an `input' in the simulations is field-like (in terms of binary fraction and separation distribution), the brown dwarf and M-dwarf binaries are usually so tight that dynamical evolution does not alter the shape of the separation distribution, and the only effect of dynamics is to slightly lower 
the initial binary fraction. On the other hand, the G-dwarf and A-star separation distributions are significantly altered, owing to the prevelance of wide ($>10^3$\,au) systems in these distributions. 

The problem then becomes one of forming a population of wide binary systems with G-dwarf and A-star primaries. This is readily achieved if part of the observed field population originates from the dissolution of supervirial star forming regions \citep{Kouwenhoven10,Moeckel10}, or the replenishment of soft binaries in 
star clusters \citep{Moeckel11a}. In the former scenario, the formation of binaries with higher mass primaries (e.g.\,\,G-dwarf and A-star) is preferred over lower mass (M-dwarf and brown dwarf) primaries, so the field binary population naturally occurs from a mix of sub-virial and supervirial SF regions (or put simply, clusters and associations). 
Whilst this is still ``fine tuning'' to some extent, the advantage of the field population over the common population scenario as an input is that the brown dwarf and M-dwarf binaries are not overproduced.

Furthermore, the argument for binary disruption rests on the assumption that the field population originates in dense star forming environments. \citet{Bressert10} suggested that only 25\,per cent of \emph{current} star forming regions are dense enough to cause destruction of binaries, although the reliance on the local surface density 
in this work was later questioned by a number of authors \citep{Moeckel12,Gieles12,Pfalzner12}. \citet{Parker12d} suggested that the fraction of `dynamically active' stars quoted in \citet{Bressert10} could be revised upward to $\sim$50\,per cent. Binaries in the field are of a similar age to the Sun (or older) and it is possible that 
star formation occured in more dense environments at earlier epochs \citep[][and references therein]{Longmore14}. However, it seems unlikely that \emph{all} field stars originate in very dense environments and a more quiescent formation environment would imply less dynamical sculpting of the field binary population. 



With these arguments in mind, we suggest that the binary populations observed in the field are most likely \emph{indicative} of the primordial populations from star formation -- and dynamical evolution has not played a significant role in altering the separation distributions or the binary fraction. \citet{Parker13b} show that the 
companion mass ratio distribution is not significantly altered by dynamical interactions and we suggest that this can be extended (with caution -- see below) to the separation distribution and binary fraction. Hydrodynamic simulations 
that predict the orbital properties and fraction of binaries \citep[e.g.][]{Donate04b,Goodwin04b,Offner10} can then be compared to the observations of the field, especially if they make predictions for the binary properties as a function of primary mass \citep{Bate12}. 

Indeed, earlier work by \citet{Moeckel10} took the end output of hydrodynamical simulations by \citet{Bate09} and evolved them for a further 10\,Myr with an $N$-body code. They also noted that very little dynamical processing occured during the subsequent $N$-body evolution, and concluded that the binary properties were mainly set during the star formation phase in the hydrodynamical calculation. Whilst the simulations presented here are purely $N$-body, and the initial conditions differ to those in the $N$-body simulations 
of \citet{Moeckel10}, our conclusions are similar.

We note that our simulations do not contain any gas; whilst this is unlikely to affect the dynamical destruction of intermediate/wide binaries in a SF region \citep[or the subsequent dynamical evolution of the region, e.g.\,][]{Offner09,Kruijssen12a}, it is possible that dynamical friction in dense gas could 
lead to the orbital decay of binary stars \citep[and subsequent merging of the component stars,][]{Stahler10}. This in turn would mean that the close binaries in the field, whilst unaffected by dynamical interactions in SF regions, may not be primordial in the sense that close separation ($<$1\,au) binaries
are preferentially destroyed (and the overall binary fraction is lowered) by the effects of gas friction, if all close binaries originate from SF regions with gas densities $> 10^5$\,cm$^{-3}$ \citep{Korntreff12}. 

However, given that a single common binary population (identical binary fraction and mean separation) across all primary masses is ruled out by dynamical interactions due to the observational constraint of decreasing mean separation and binary fraction with primary mass, we appeal to Occam's Razor 
and suggest that as the field is a better input population to fit the observations after dynamical evolution, it is likely to be closer to the primordial population than most other distributions.

Even if dynamical evolution has only had a modest effect on the binary fraction and orbital parameters of most systems in the field, we emphasize that much work still remains to be done in understanding the formation and evolution of binary stars \citep[e.g.][]{Reipurth14}. Most observations of 
young star forming regions can only probe binaries with separations in the intermediate regime \citep[10s -- 1000s\,au -- ][]{King12a,King12b,Duchene13b}. It is these systems that are most likely to change through 
dynamical interactions \citep{Heggie75,Hills75a,Hills75b} -- and often in an unpredictable way \citep{Parker12b}. It is quite possible that dynamical interactions may play a significant role in altering the binary properties in some regions -- especially low-mass regions where the effects of dynamics can be highly stochastic \citep{Becker13}. 
We also have very little information on `hard' binaries in clusters -- which are likely to influence the global dynamical evolution of star clusters and star-forming regions \citep[e.g.][]{Allison11,Geller13b,Parker14b}. 
 

Furthermore, observations of Class 0 protostars by \citet{Chen13} suggest that the binary fraction of these objects is higher than that found in the later-stage Class~I Young Stellar Objects (YSOs) by \citet{Connelley08}, which in turn is higher than the fraction found for G-dwarf Main Sequence binaries by \citet{Raghavan10}. \citet{Connelley08} also 
show that the shape of the spearation distribution of YSO binaries is not log-normal, but rather flat in log-space [an \citet{Opik24} distribution]. We first note that the difficulty in determining accurate masses for Class~0/Class~I objects means that the binary poperties of the objects in the samples of \citet{Connelley08} and \citet{Chen13} may 
not always be comparable with the \citet{Raghavan10} sample, as their eventual MS primary masses may be different. However, regardless of this caveat, we have shown in this paper that external perturbations on binaries from dynamical interactions in SF regions are unlikely to account for a drastic change in the binary fraction and orbital 
separation distribution of binary systems. Therefore, if the binary fraction and separation distribution of protostellar and YSO binaries are different to that of MS binaries, it is possible that the differences may be due to internal dynamical evolution of these young systems \citep{Reipurth14}, rather than external dynamical evolution in the SF 
environment. This is particularly relevant as a high fraction of stars form in triple, quadruple and higher order multiple systems \citep{Tokovinin08,Tokovinin14}, which may become (or even form) dynamically unstable \citep[][and references therein]{Reipurth12,Reipurth14}.

\section{Conclusions}
\label{conclude}

We have presented $N$-body simulations of the evolution of dense star-forming regions to determine the impact of dynamical interactions on different primordial binary populations. Our results are summarised as follows:\\

(i) The field binary population is not the end-product of dynamical processing of a common primordial binary population \citep{Kroupa95b} with an initial binary fraction of unity and an excess of systems with intermediate/wide separations ($100 - 10^4$\,au). 
Whilst dynamical evolution does cause the binary fraction to decrease as a function of primary mass with these initial conditions, the binary fractions are typically too high and the difference in mean separation as a function of decreasing primary mass as observed 
for binaries in the Galactic field \citep{Duchene13b} is not reproduced. 

(ii) If the primordial binary population is similar to that in the field, very few brown dwarf and M-dwarf binaries are destroyed, due to their average semi-major axis being well-within the `hard' binary regime at 4.6\,au \citep[the brown dwarfs;][]{Burgasser07} and 16\,au 
\citep[the M-dwarfs;][]{Bergfors10,Janson12}. This suggests that the majority of binaries have been unaffected by dynamical interactions, as M-dwarfs make up the majority of stars in the Universe \citep{Bastian10,Bochanski10}. 

(iii) G-dwarf and A-star binaries in the field are observed to have wider separations, with peaks at 50\,au \citep{Raghavan10} and 389\,au \citep{DeRosa14}, respectively. Whilst some intermediate/wide G-dwarf and A-star binaries could be destroyed in dense star forming environments, 
the formation of wide binaries during the dissolution of star-forming regions \citep{Kouwenhoven10,Moeckel10} is a strong function of primary mass. Therefore, if the field is a mixture of systems from clusters and associations (or just from expanding associations) then the formation 
of these observed wide binary G-dwarf and A-star systems are a natural outcome of the dissolution of star-forming regions into the field.

(iv) The combination of points (ii) and (iii), and the possibility that not all star forming regions are dense enough to dynamically affect binaries \citep{Bressert10}, leads us to suggest that the binary fractions and semi-major axis distributions in the field are \emph{indicative} of the primordial population.

(v) However, more complete observations of pre-main sequence binaries in star forming regions are desperately required in order to determine the origin of the Galactic field population. Currently, observed visual binaries straddle the hard/soft boundary, 
and it is these systems that are most likely not to be indicative of the primordial population, especially as they are susceptible to stochastic destruction \citep{Parker12b}.

The above conclusions do not detract from the fact that binaries are an important part of the star formation process, and in principle can tell us much about star formation. All we have shown here is that their primordial orbital separations and overall fraction 
are probably not drastically altered in dense star-forming environments, and the need for star formation to produce a large excess of primordial wide-intermediate binary systems to compensate for dynamical destruction in star-forming regions is not required. 

Finally, we note that our work is not actually in conflict with earlier numerical simulations which did require a significant degree of processing of (mainly) intermediate M-dwarf binaries to explain the field population \citep{Kroupa95b}; this earlier work did not have the updated binary statistics on brown dwarfs, M-dwarfs and A-stars at its disposal.

\section*{Acknowledgements}

We thank the anonymous referee for their comments and suggestions, which improved the original manuscript. 
RJP acknowledges support from the Royal Astronomical Society in the form of a research fellowship. MRM acknowledges support from the Swiss National Science Foundation (SNF). The simulations in this work were performed on the \texttt{BRUTUS} computing cluster at ETH Z{\"u}rich.

\bibliographystyle{mn2e}
\bibliography{general_ref}

\label{lastpage}

\end{document}